\documentclass[twocolumn,superscriptaddress,amsmath,amssymb,aps,pra]{revtex4}
\usepackage{amsmath,amsthm,amssymb}
\usepackage{latexsym}
\usepackage{amscd,graphicx}
\usepackage[titletoc]{appendix}
\usepackage{epsfig}
\usepackage{epstopdf}
\usepackage[normalem]{ulem}
\usepackage[dvipsnames]{xcolor}
\usepackage[colorlinks=true,linkcolor=red,citecolor=blue]{hyperref}

\begin{document}

\title{Higher-order exceptional point in a blue-detuned non-Hermitian cavity optomechanical system}

\author{Wei Xiong}
\altaffiliation{xiongweiphys@wzu.edu.cn}

\affiliation{Department of Physics, Wenzhou University, Zhejiang 325035, China}

\author{Zhuanxia Li}
\affiliation{Department of Physics, Wenzhou University, Zhejiang 325035, China}

\author{Guo-Qiang Zhang}
\affiliation{School of Physics, Hangzhou Normal University, Hangzhou 311121, China}

\author{Mingfeng Wang}

\affiliation{Department of Physics, Wenzhou University, Zhejiang 325035, China}

\author{Hai-Chao Li}

\altaffiliation{hcl2007@foxmail.com}
\affiliation{College of Physics and Electronic Science, Hubei Normal University, Huangshi 435002, China}

\author{Xiao-Qing Luo}

\altaffiliation{xqluophys@gmail.com}
\affiliation{Hunan Province Key Laboratory for Ultra-Fast Micro/Nano Technology and Advanced Laser Manufacture,
	School of Electrical Engineering, University of South China, Hengyang 421001, China}

\author{Jiaojiao Chen}
\altaffiliation{jjchenphys@hotmail.com}
\affiliation{School of Physics and Optoelectronics Engineering, Anhui University, Hefei 230601, China}
\affiliation{Hefei Preschool Education College, Hefei 230013, China}

\date{\today }

\begin{abstract}
Higher-order exceptional points (EPs) in non-Hermitian systems have attracted great interest due to their advantages in sensitive enhancement and distinct topological features. However, realization of such EPs is still challenged because more fine-tuning parameters is generically required in quantum systems, compared to the second-order EP (EP2). Here, we propose a non-Hermitian three-mode optomechanical system in the blue-sideband regime for predicting the third-order EP (EP3). By deriving the pseudo-Hermitian condition for the proposed system, one cavity with loss and the other one with gain must be required. Then we show EP3 or EP2 can be observed when the mechanical resonator (MR) is neutral, loss or gain. For the neutral MR, we find both two degenerate or two non-degenerate EP3s can be predicted by tuning system parameters in the parameter space, while four non-degenerate EP2s can be observed when the system parameters {deviate} from EP3s, which is distinguished from the previous study in the red-detuned optomechanical system. For the gain (loss) MR, we find only two degenerate EP3s or EP2s can be predicted by tuning enhanced coupling strength. Our proposal provides a potential way to predict higher-order EPs or multiple EP2s and study multimode quantum squeezing around EPs using the blue-detuned non-Hermitian optomechanical systems. 
\end{abstract}


\maketitle

\section{introduction}
Cavity optomechanical (COM) systems,  emerged as a promising platform in quantum information science, have been paid considerable attention both theoretically and experimentally~\cite{Aspelmeyer}. The simplest COM system is made up of a mechanical resonator (MR) nonlinearly coupled to a cavity via radiation pressure, which can be well controlled by strong driving fields.  In such mystical systems, abundant effects including sensing~\cite{Schreppler-2014,Wu-2017,Gil-Santos-2020,Fischer-2019}, ground-state cooling~\cite{Chan-2011,Teufel-2011}, squeezed light generation~\cite{Purdy-2014,Safavi-Naeini-2013,Aggarwal-2020}, nonreciprocal transport~\cite{Xu-2019,Shen-2016}, optomechanically induced transparency~\cite{Kronwald-2013,Weis-2010,Liuy-2013}, coupling enhancement~\cite{Xiong-2021,Lu-2013,Xiong2-2021}, and nonlinear behaviors ({e.g., bi- and tristability and chaos} )~\cite{Lu-2015,Xiong3-2016} have been investigated.

In addition, COM systems have shown huge potential in studying exceptional points (EPs) of non-Hermitian systems~\cite{Xiongwei-2021,Jing-2014,Xu-2016,LYL-2017,ZhangJQ-2021,XuH-2021,Xuxw-2015}, at which both eigenvalues and eigenvectors coalesce. This is due to the fact that practical COM systems can be characterized by effective non-Hermitian Hamiltonians when decoherence arising from surrounding {environment} is considered. Moreover, the driven COM systems can provide fine-tuning parameters for requirement of realizing EPs, assisted by strong driving fields. Owing to these, EPs have been intensively studied in COM systems, especially for the second-order EPs (EP2s)~\cite{Jing-2014,Xu-2016,LYL-2017,ZhangJQ-2021,XuH-2021,Xuxw-2015} where two eigenvalues and the corresponding eigenvectors coalesce~\cite{Minganti-2019,Zhang-2021,Ozdemir-2019,Mostafazadeh1-2002,Mostafazadeh2-2002,Konotop-2016,Bender-2013,Parto-2021,Bergholtz-2021,Wiersig-2020,Feng-2017,El-Ganainy-2018}. Besides, EP2s are also studied in other systems~~\cite{Doppler-2016, Zhang-2017,Harder-2017,Quijandria-2018,Zhang-2019-2,Naghiloo-2019,ZhangGQ}. Around EP2s, lots of fascinating phenomena like unidirectional invisibility~\cite{Peng-2014,Lin-2011,Chang-2014}, single-mode lasing~\cite{Feng-2014,Hodaei-2014}, sensitivity enhancement~\cite{Chen-2017,Hokmabadi-2019}, energy harvesting~\cite{Fern-2021}, protecting the classification of exceptional nodal topologies~\cite{Marcus-2021}, electromagnetically induced transparency~\cite{Guo-2009,Wang-2019,Wang1-2020,Lu-2021}, {and quantum squeezing~\cite{Miranowicz-2019,Mukherjee-2019,Luo-2022} can be studied}. 

Instead of EP2s, non-Hermitian systems can also host higher-order EPs, where more than two
eigenmodes coalesce into one~\cite{Graefe-2008,Heiss-2008,Demange-2012,Heiss-2016,Jing-2017,Ge-2015,Lin-2016,Quiroz-2019,Bian-2020}. It has been shown that higher-order EPs can exhibit greater advantages than EP2s in spontaneous emission enhancement~\cite{Lin-2016}, sensitive detection~\cite{Hodaei-2017,Zeng-2021,Wang-2021,Zeng-2019}, topological characteristics~\cite{Ding-2016,Delplace-2021,Mandal-2021}. With these superiorities, higher-order EPs are being intensively studied in various systems~\cite{Roy-2021,Zhong-2020,Zhang-2020,Pan-2019,Zhang-2019,Kullig-2019,Kullig-2018,Schnabel-2017,Nada-2017,Wang-2020} but attract less attention in non-Hermitian COM systems. For this, how to construct higher-order EPs in non-Hermitian COM systems is strongly demanded.

We also note that EPs in non-Hermitian COM systems, including EP2s and EP3s, are mainly focused in the red-sideband regime~\cite{Xiongwei-2021,Jing-2014}. In this regime, fast oscillating terms related to mode squeezing are neglected. This limits nonclassical quantum effects such as quantum squeezing investigation around EPs. For this, we here theoretically propose a {paradigmatic} COM system consisting of a MR coupled to two cavities via radiation pressure for predicting EP3s, where two cavities are respectively passive (loss) and active (gain), and driven by two blue-detuned classical fields. First, we derive an effective non-Hermitian Hamiltonian for the proposed COM system and analytically give the pseudo-Hermitian condition of the proposed COM system in the general case. Then, three scenarios are specifically considered in the pseudo-Hermitian condition: (i) the neutral MR; (ii) the passive MR; (iii) the active MR.
In case (i), the proposed non-Hermitian COM system with symmetric coupling strength can host both two degenerate EP3s and two non-degenerate EP3s in the parameter space. When we tune the system parameters {deviation} from the critical paraters at EP3s, four non-degenerate EP2s can be predicted, which is different from the situation in the red-sideband non-Hermitian COM system. For the cases (ii) and (iii), the proposed non-Hermitian COM system is required to have asymmetric coupling strength for satisfying pseudo-Hermitian condition. We find only two degenerate EP3s or two degenerate EP2s can be predicted. By investigating the effects of system paramters on EP3s or EP2s, we find large coupling strength or large frequency detuning is benefit to observe EPs more clearly.
Our proposal provides a promising path to study nonclassical quantum effects around EP2s and EP3s in non-Hermitian COM systems, and it is the first scheme to study higher-order EPs in the blue-detuned COM system, {although two-mode quantum squeezing has been investigated in a system with pseudo-anti-parity-time symmetry~\cite{Luo-2022}}. 

 This paper is organized as follows. In Sec. II, the model is described and the system effective Hamiltonian is given.  Then we derive the pesudo-Hermitian condition for the considered non-Hermitian COM system in Sec. III. In Sec. IV, critical parameters of the proposed COM system at EP3 are anatically derived. In Sec. V, phase diagram of the descriminant for the characteristic equation is studied to predict EP3 and EP2. In Sec. VI, EP3 and EP2 in three cases are specifically studied. Finally, a conclusion is given in Sec. VII.

\section{Model and Hamiltonian}

\begin{figure}
	\includegraphics[scale=0.35]{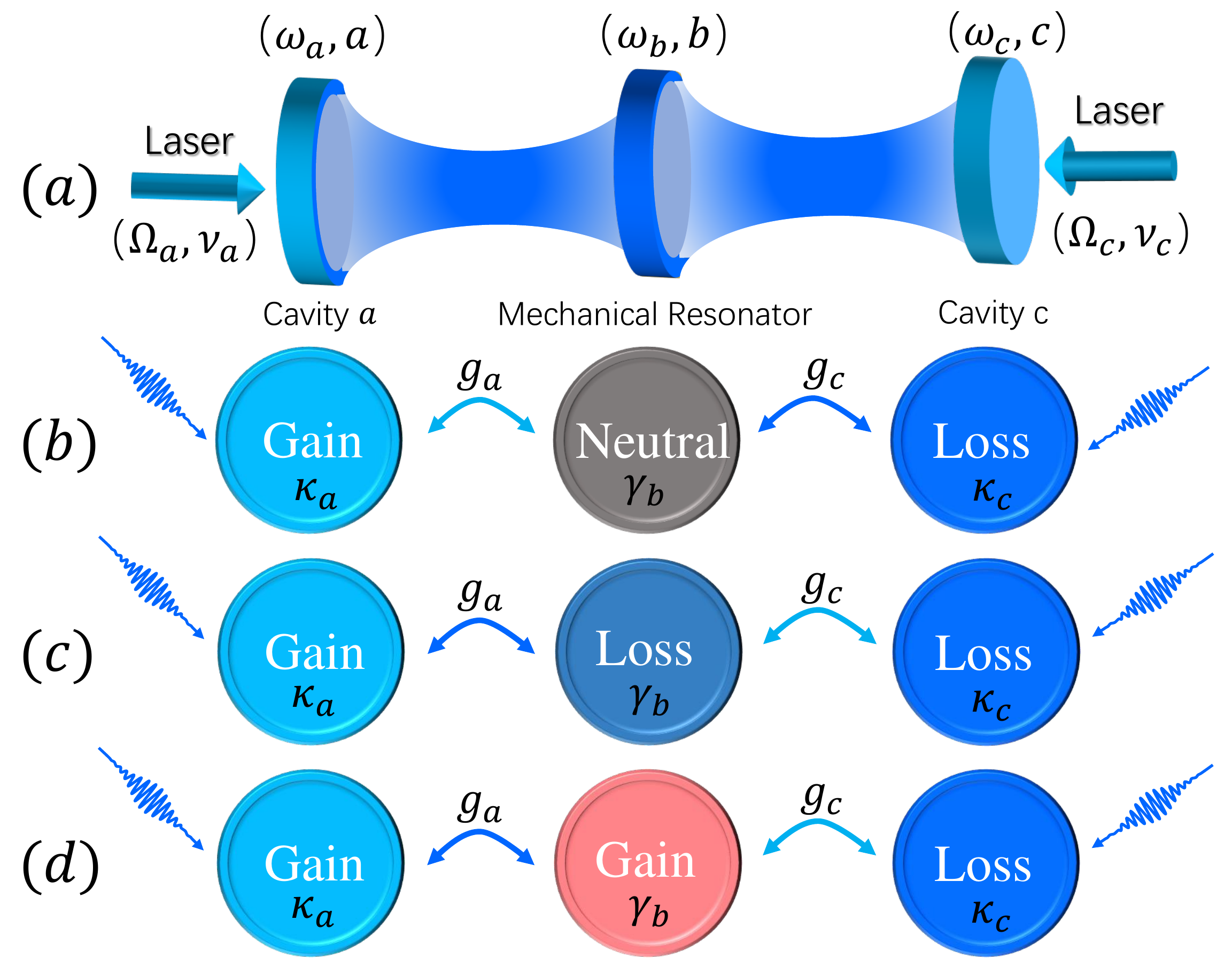}
	\caption{(a) Schematic diagram of the proposed blue-detuned three-mode optomechanical system. It consists of two cavities labeled as cavity $a$ and cavity $c$, with respective frequencies $\omega_a$ and $\omega_c$, coupled to a common MR with frequency $\omega_b$. The two cavities are driven by two blue-detuned laser fields with frequencies $\nu_a$ and $\nu_b$. The corresponding amplitudes are $\Omega_a$ and $\Omega_c$. (b) The neutral MR, $\gamma_b=0$, is considered. (c) The loss MR, i.e., $\gamma_b>0$, is considered. (d) The gain MR, $\gamma_b<0$, is considered. In (a), (b) and (c), $\kappa_a(<0)$ is the gain rate of the cavity $a$, $\kappa_c(>0)$ is the loss rate of the cavity. $g_{a(c)}$ is the single-photon optomechanical coupling strength of the MR coupled to the cavity $a~(c)$.}\label{fig1}
\end{figure}
We consider an experimental three-mode optomechanical system~\cite{Dong-2012,Hill-2012,Andrews-2014} consisting of two driven cavities (labeled as cavity $a$ and cavity $c$) coupled to a MR with frequency $\omega_b$ via radiation pressure (see Fig.~\ref{fig1}). At the rotating frame respect to two laser fields, the Hamiltonian of the total system can be written as (setting $\hbar=1$)~\cite{Zhangk-2015}
\begin{align}
	H_{\rm total}=&\delta_a a^\dag a+\omega_b b^\dag b+\delta_c c^\dag c\notag\\
	&+g_aa^\dag a(b^\dag+b)+g_cc^\dag c(b^\dag+b)+H_D,\label{eq1}
\end{align}
where $\delta_{a(c)}=\omega_{a(c)}-\nu_{a(c)}$, with $\omega_{a(c)}$ being the frequency of the cavity $a~(c)$ and $\nu_{a(c)}$ the frequency of the laser field acting on the cavity $a~(c)$, is the frequency detuning of the cavity $a~(c)$ from the laser field acting on the cavity $a~(c)$.  $g_a$ and $g_c$ are the single-photon optomechanical coupling strengths of the MR coupled to the cavity $a$ and cavity $c$, respectively. The operators $a~(c)$ and $a^\dag~(c^\dag)$ are the annihilation and creation operators of the cavity $a~(c)$. The last term $H_D=i(\Omega_a a^\dag+\Omega_c c^\dag)+{\rm H.c.}$ in Eq.~(\ref{eq1}) represents the coupling between two cavities and two laser fields with Rabi frequencies $\Omega_a$ and $\Omega_c$. As we are interested in the blue-sideband regime of the proposed COM system, thus $\delta_a,~\delta_c<0$ is assumed below.

{In the strong-field limit, the nonlinear COM system can be linearized by writing each operator as the expectation value ($a_s,~b_s,~c_s$) plus the corresponding fluctuation ($\delta a,~\delta_b,~\delta_c$). Neglecting the higher-order fluctuations, the linearized Hamiltonian including dissipations can be given by (see details in the Appendix)}

\begin{align}
	H_{\rm eff}=&(\delta_a^\prime-i\kappa_a) \delta a^\dag  \delta a+(\omega_b-i\gamma_b) \delta b^\dag \delta b+(\delta_c^\prime-i\kappa_c) \delta c^\dag \delta c\notag\\
	&+G_a (\delta a^\dag \delta b^\dag+\delta a \delta b)+G_c(\delta b^\dag\delta c^\dag +\delta b\delta c ).\label{eq16}
\end{align}
{Here, fast oscillating terms have been discarded with the condition $|\delta_{a(c)}^\prime+\omega_b|\ll|\delta_{a(c)}^\prime-\omega_b|$ and $|G_{a(c)}|\ll|\delta_{a(c)}^\prime|$, where $\delta_{a(c)}^\prime$ is the effective frequency detuning of the cavity $a~(c)$ shifted by the displacement of the mechanical resonator, and $G_{a(c)}$ is the effective optomechanical coupling strength enhanced by the photon number in the cavity $a~(c)$.} This effective Hamiltonian is the typical three-mode squeezing Hamiltonian without dissipations. For convenience, we assume $G_a$ and $G_c$ to be real, which can be realized by tuning the phase of two laser fields.

\section{Pseudo-Hermitian condition}
The effective Hamiltonian in Eq.~(\ref{eq16}) can also be equivalently expressed as
\begin{equation}
H_{\rm eff}=\left(
\begin{array}{ccc}
\delta_a^\prime-i\kappa_a & G_a & 0\\
-G_a^* & -\omega_b-i\gamma_b & -G_c^*\\
0 &  G_c  & \delta_c^\prime-i\kappa_c
\end{array}
\right)\label{eq17}
\end{equation}
is just the matrix form of $H_{\rm eff}$.
For the non-Hermitian Hamiltonian in Eq.~(\ref{eq17}), three eigenvalues can be predicted. When these three eigenvalues are all real, or one is real and the other
two are a complex-conjugate pair, the considered three-mode optomechanical system characterized
by the Hamiltonian in Eq.~(\ref{eq16}) or Eq.~(\ref{eq17}) is pseudo-Hermitian~\cite{Mostafazadeh1-2002,Mostafazadeh2-2002}. For the pseudo-Hermitian systems, 
the characteristic polynomial equation
\begin{align}
	|H_{\rm eff}-\Omega \mathbb{I}|=0\label{eq18}
\end{align}
is the same as
\begin{align}
	|H_{\rm eff}^*-\Omega \mathbb{I}|=0,\label{eq19}
\end{align}
where $H_{\rm eff}^*$ is the complex conjugate transpose of $H_{\rm eff}$, $\mathbb{I}$ is a $3\times3$ identity matrix, and $\Omega$ denotes the eigenvalue of the effective Hamiltonian $H_{\rm eff}$. By expanding Eqs.~(\ref{eq18}) and (\ref{eq19}), and comparing the corresponding coefficients, we can obtain
\begin{align}\label{eq20}
       \kappa_a+\gamma_b+\kappa_c=&0,\notag\\
	\gamma_b(\delta_a^\prime+\delta_c^\prime)=&\kappa_a(\omega_b-\delta_c^\prime)+\kappa_c(\omega_b-\delta_a^\prime),\\
	(\delta_a^\prime\omega_b+\kappa_a\gamma_b)\kappa_c=&G_a^2\kappa_c+G_c^2\kappa_a+(\delta_a^\prime\gamma_b-\kappa_a\omega_b)\delta_c^\prime.\notag
\end{align} 
By setting
\begin{align}
	\eta=&\kappa_a/\kappa_c,~~~\lambda=G_c/G_a,~~
	\Delta_{\rm a(c)}=\delta_{a(c)}^\prime+\omega_b,
\end{align}
Eq.~(\ref{eq20}) can be further simplified as
\begin{align}\label{phc}
\gamma_b+(1+\eta)\kappa_c=0,\notag\\
\Delta_c+\Delta_a\eta=0,\\
(1+\lambda^2\eta)G_a^2+\eta(1+\eta)(\Delta_a^2+\kappa_c^2)=0.\notag
\end{align}
Obviously, only when the conditions in Eq.~(\ref{phc}) are simutaneously satisfied, the considered three-mode optomechanical system is pseudo-Hermitian. From the first condition in Eq.~(\ref{phc}), we can see that the decay rates from the cavity $a$, the mecahanical resonator and the cavity $c$ are required to be balanced. This means gain effect must be introduced to the considered system. From the third condition, $\eta<0$ is obtained, which shows one loss cavity and the other gain cavity are always needed to satisfy the pseudo-Hermitian condition for the proposed COM system in the blue-sideband regime. This situation is completely different from the previous study of EPs using a COM system in the red-sideband regime. Without loss of generality, the cavity $a$ with gain, and the cavity $c$ with loss are taken, i.e., $\kappa_a<0$ and $\kappa_c>0$. From the third equation in Eq.~(\ref{phc}), it is not difficult to find the fact that $\lambda=1$ when $\eta=-1$, which indicates that the coupling strengths between the MR and two cavities must be uniform, i.e., $G_a=G_c$. When $\eta\neq-1$, 
\begin{align}
	(1+\eta)(1+\lambda^2\eta)>0 ~~{\rm for}~~\eta\neq-1\label{eq22}
\end{align}
is directly given by the third equality in Eq.~(\ref{phc}), which in turn gives rise to a boundary for the parameter $\lambda$ or equivalently $G_a$ and $G_c$. Such the boundary can be achieved here owing to the tunable parameters $\Delta_a,~\Delta_c,~G_a$ and $G_c$. 

\section{critical parameters at EP3}
When the pseudo-Hermitian conditions in Eq.~(\ref{phc}) are satisfied and  $x=\Omega+\omega_b$ is defined, the characteristic equation in Eq.~(\ref{eq18}) reduces to
\begin{align}
	x^3+c_2 x^2+c_1 x+c_0=0,\label{cubic}
\end{align}
where
\begin{align}
	c_2=&(\eta-1)\Delta_a,\notag\\
	c_1=&(1+\lambda^2)G_a^2-\eta\Delta_a^2+(1+\eta+\eta^2)\kappa_c^2,\\
	c_0=&(\eta-\lambda^2)G_a^2\Delta_a-(1+\eta)^2(1-\eta)\kappa_c^2\Delta_a.\notag
\end{align}\label{coe}
According to Cardano’s formula~\cite{kORN-1968}, the solutions of this characteristic equation is determined by the discriminant
\begin{align}
	D=B^2-4A C,\label{eq25}
\end{align}
where
\begin{align}
	A=&c_2^2-3c_1,~~B=c_1c_2-9c_0,~~C=c_1^2-3c_0 c_2.\label{eq26}
\end{align}
For $D<0$, Eq.~(\ref{cubic}) has three real roots. But for $D>0$, Eq.~(\ref{cubic}) only has one real root and the other two are complex roots. Interestingly,  three roots coalesce to the same value $\Omega_{\rm EP3}$ at $D=0$ with $A=B=0$, which is so-called EP3. For the case of $D=0$ but $A\neq0$ and $B\neq0$, only two roots coalesce to the value $\Omega_{\rm EP2}$, corresponding to EP2.
\begin{figure}
	\includegraphics[scale=0.5]{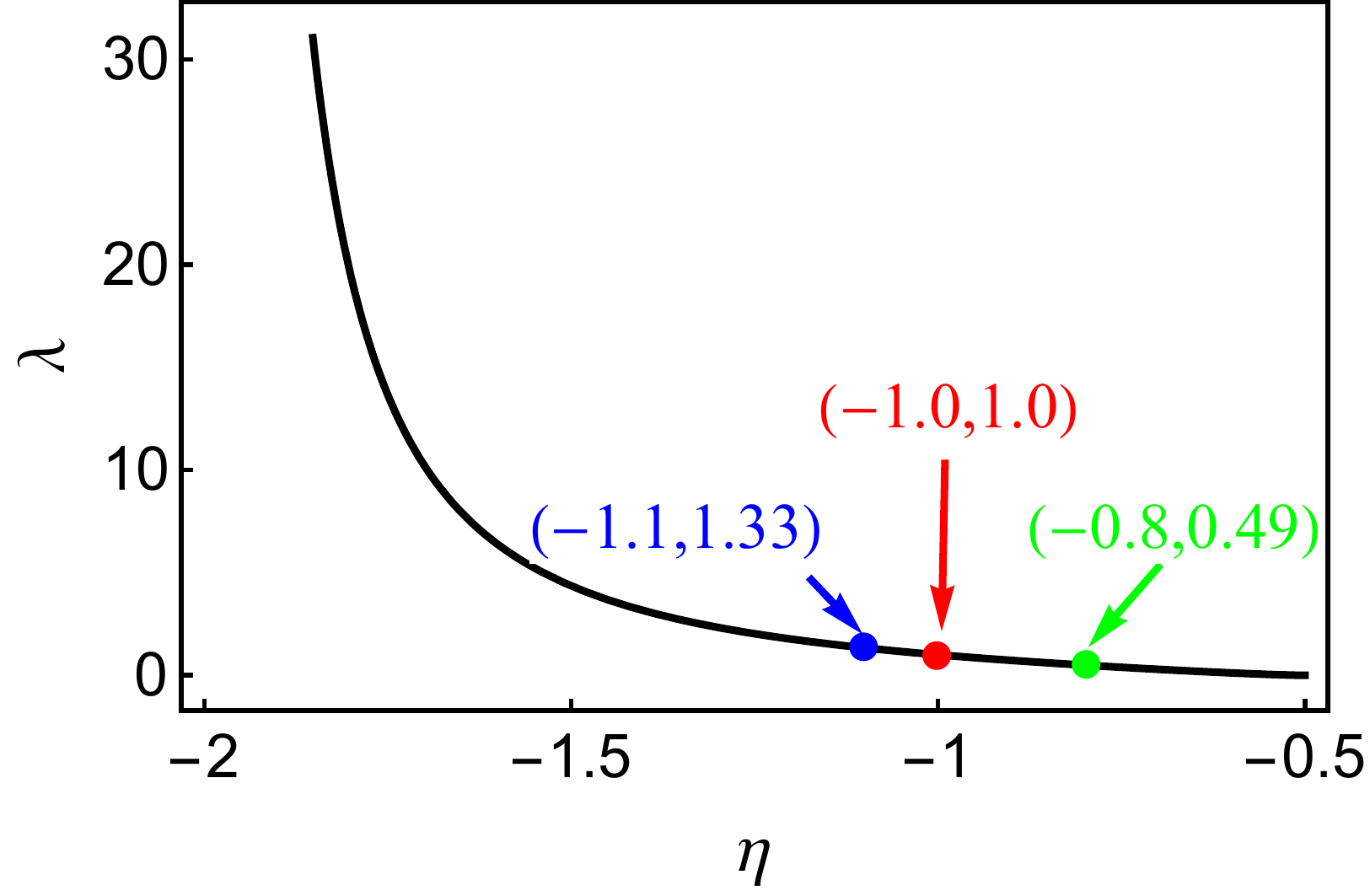}
	\caption{The parameter $\lambda$ at EP3 as a function of $\eta$. The red, blue and green dots respectively denote $(\eta,\lambda_{\rm EP3})=(-1,1), (-1.1,1.33), (-0.8,0.49)$. }\label{fig1_1}
\end{figure}
\begin{figure*}
	\includegraphics[scale=0.348]{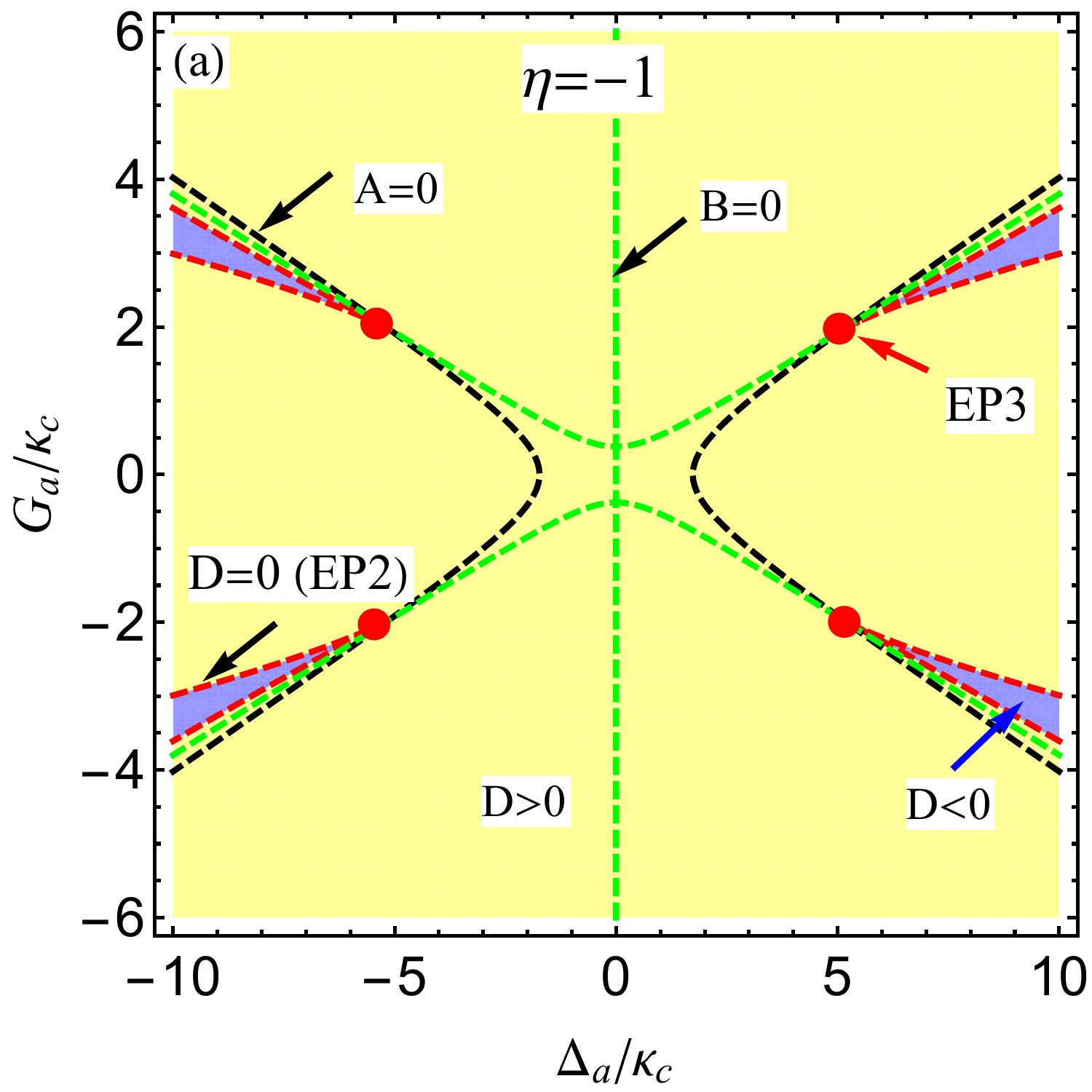}
	\includegraphics[scale=0.305]{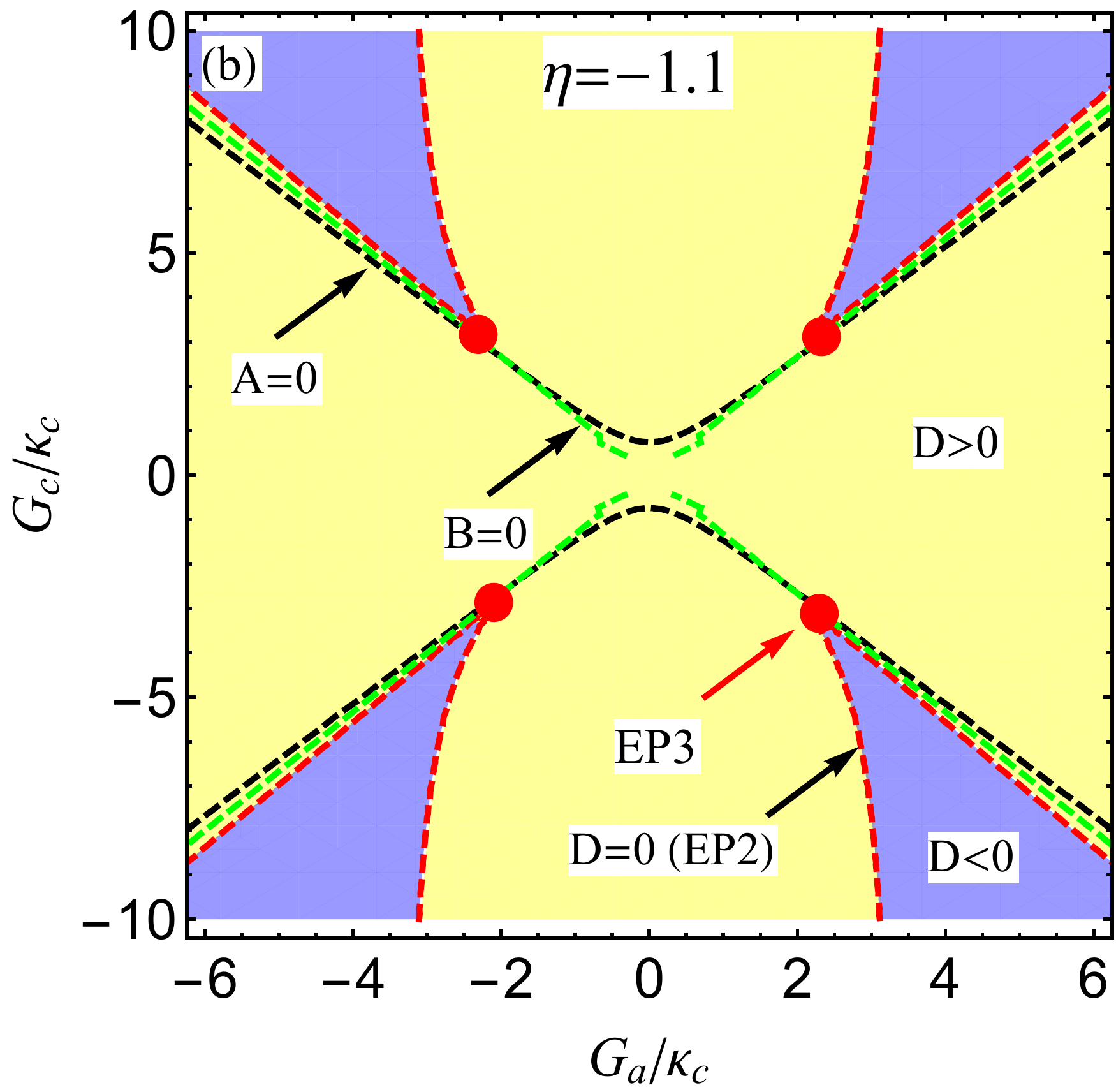}
	\includegraphics[scale=0.3]{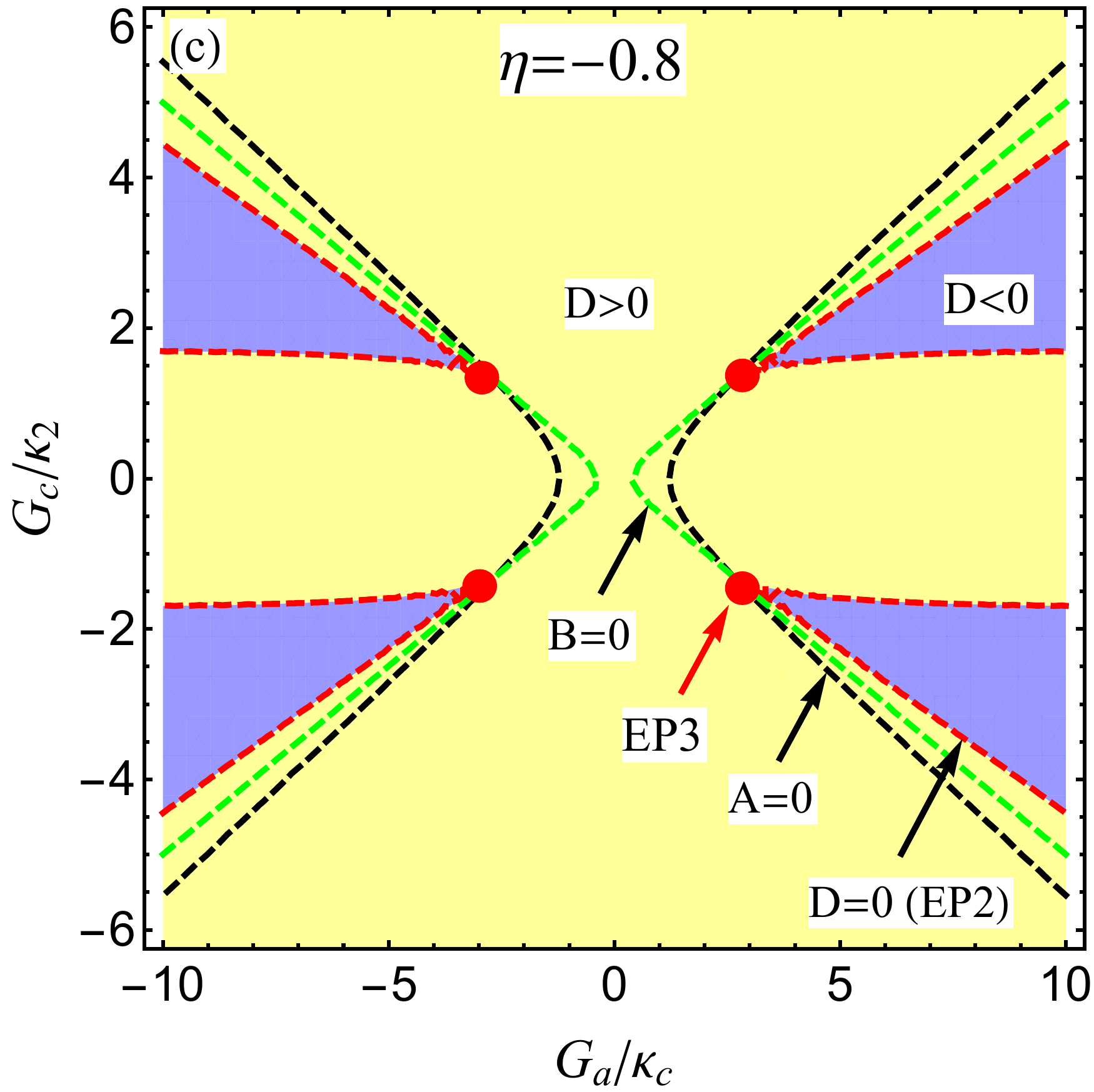}
	\caption{The phase diagram of the discriminant given by
		Eq.~(\ref{eq25}) vs the normalized parameters (a) $G_a/\kappa_c$ and $\Delta_a/\kappa_c$; (b) and (c) $G_a/\kappa_c$ and $G_c/\kappa_c$. In (a), (b) and (c), $\eta$ is respectively taken as $\eta=-1$, $\eta=-1.1$, and $\eta=-0.8$.}\label{fig2}
\end{figure*}

Below we analytically derive the critical parameters at EP3. When EP3 appears at $\Omega=\Omega_{\rm EP3}$, we have
\begin{align}
	(\Omega-\Omega_{\rm EP3})^3 = 0
\end{align}
Comparing coefficients of this equation with Eq.~(\ref{cubic}),
\begin{align}
	-3x_{\rm EP3}=c_2,~~3x_{\rm EP3}^2=c_1,~~x_{\rm EP3}^3=-c_0\label{eq}
\end{align}
are obtained. The first equation directly leads to
\begin{align}
	x_{\rm EP3}=\frac{1}{3}(1-\eta)\Delta_a.\label{s1}
\end{align}
Substituting this solution back into the second equation in Eq.~(\ref{eq}), the critical coupling strength at EP3 is given by
\begin{align}
	G_{\rm a, EP3}=&2\kappa_c\left[-\frac{3(1+\lambda^2)}{1+\eta+\eta^2}-\frac{1+\lambda^2\eta}{\eta(1+\eta)}\right]^{-1/2},\label{eq31}
\end{align}
where 
\begin{align}
	\Delta_{\rm a,EP3}=\pm\left[-\frac{1+\lambda^2\eta}{\eta(1+\eta)}G_{\rm a, EP3}^2-\kappa_c^2\right]^{1/2}
\end{align}
is derived from the third equation in Eq.~(\ref{phc}). As $\Delta_{\rm a,EP3}^2\geq0$, so the minimal value of $G_a$ for predicting EP3 is 
\begin{align}
	G_{\rm a,EP3}^{\rm min}=\left[-\frac{\eta(1+\eta)}{1+\lambda^2\eta}\kappa_c\right]^{1/2}.
\end{align}
At EP3, the parameter $\lambda$ is required to meet
\begin{align}
\lambda_{\rm EP3}=&\left[\frac{2\eta+1}{\eta(\eta+2)}\right]^{3/2},\label{eq34}
\end{align}
which is obtained by substituting the solution in Eq.~(\ref{s1}) back into the third equality in Eq.~(\ref{eq}). To see the dependent relationship between $\lambda$ and $\eta$ at EP3 more clearly, we plot $\lambda$ as a function of $\eta$ in Fig.~\ref{fig1_1}. Obviously, $\lambda$ monotonously decreases with the absolute value of $\eta$. Eq.~(\ref{eq34}) also requires $\eta$ to satisfy
\begin{align}
	(\eta+2)(2\eta+1)<0.\label{eq35}
\end{align}
Combine Eqs.~(\ref{eq22}) and (\ref{eq35}) together,  the parameter $\eta$ for predicting EP3 can take 
\begin{equation}
\left\{
\begin{aligned}
-2<&\eta<-1,\\
&\eta=-1,\\
-1<&\eta<-\frac{1}{2},
\end{aligned}
\right.\label{eq36}
\end{equation}
leading to
\begin{equation}
\left\{
\begin{aligned}
&\gamma_b>0,~~~~~{\rm loss~mechanical~resonator},\\
&\gamma_b=0,~~~~~{\rm neural~mechanical~resonator},\\
&\gamma_b<0,~~~~~{\rm gain~mechanical~resonator}.
\end{aligned}
\right.\label{eq37}
\end{equation}
The corresponding value of $\lambda_{\rm EP3}$ is given by Eq.~(\ref{eq34}).

\section{Phase diagram for prediction of EP3 and EP2}
Next, we numerically predict  EP3 and EP2 via phase diagram of the discriminant [see Eq.~(\ref{eq25})] with in the three cases given by Eq.~(\ref{eq36}).
\subsection{$\eta=-1$}
When $\eta=-1$, i.e., $\kappa_a=-\kappa_c$, two optomechanical cavities are gain-loss balanced, Eq.~(\ref{phc}) reduces to
\begin{align}
\gamma_b=0,~~\Delta_{\rm c}=\Delta_{\rm a},~~\lambda_{\rm EP3}=1.
\end{align}
For the condition $\gamma_b=0$, it is difficult to be perfectly satisfied. But for COM systems, the decay rate of the MR is in general much smaller than the decay rate of the optomechanical cavity, i.e., $\gamma_b\ll\kappa_c$. Therefore, we can safely ignore the effect of the decay rate of the MR on EP3, and thus we assume $\gamma_b\approx0$. In addition, the coefficients in Eq.~(\ref{coe}) are simplified as
\begin{align}
c_2=&-2\Delta_a,~
c_1=2G_a^2+\Delta_a^2+\kappa_c^2,~
c_0=-2G_a^2\Delta_a.
\end{align}
Correspondingly, the discriminant in Eq.~(\ref{eq25}) becomes
\begin{align}
D=\kappa_c^2\Delta_a^4-(G_a^4+10\kappa_c^2G_a^2-2\kappa_c^4)\Delta_a^2+(2G_a^2+\kappa_c^2)^3,\label{eq38}
\end{align}
and $A,~B,~C$ in Eq.~(\ref{eq26}) are
\begin{align}
A=&\Delta_a^2-6G_a^2-3\kappa_c^2,\notag\\
B=&2\Delta_a(7G_a^2-\Delta_a^2-\kappa_c^2),\\
C=&(2G_a^2+\Delta_a^2+\kappa_c^2)^2-12G_a^2\Delta_a^2.\notag
\end{align}
In Fig.~\ref{fig2}(a), we plot the phase diagram determined by the sign of the discriminant [see Eq.~(\ref{eq38})] vs the normalized parameters $\Delta_a/\kappa_c$ and $G_a/\kappa_c$, where the purple (yellow) region indicates $D>0~(D<0)$. The boundary curve in red means $D=0$. The curves in black and green respectively denote $A=0$ and $B=0$. Obviously, three curves have four cross points, that is, four EP3s in the parameter space can be found according to the Cardano's formula~\cite{kORN-1968}. Also, EP2 can be predicted by the red curve only (i.e., $D=0$, but $A\neq0$ and $B\neq0$).
\begin{figure*}
	\includegraphics[scale=0.38]{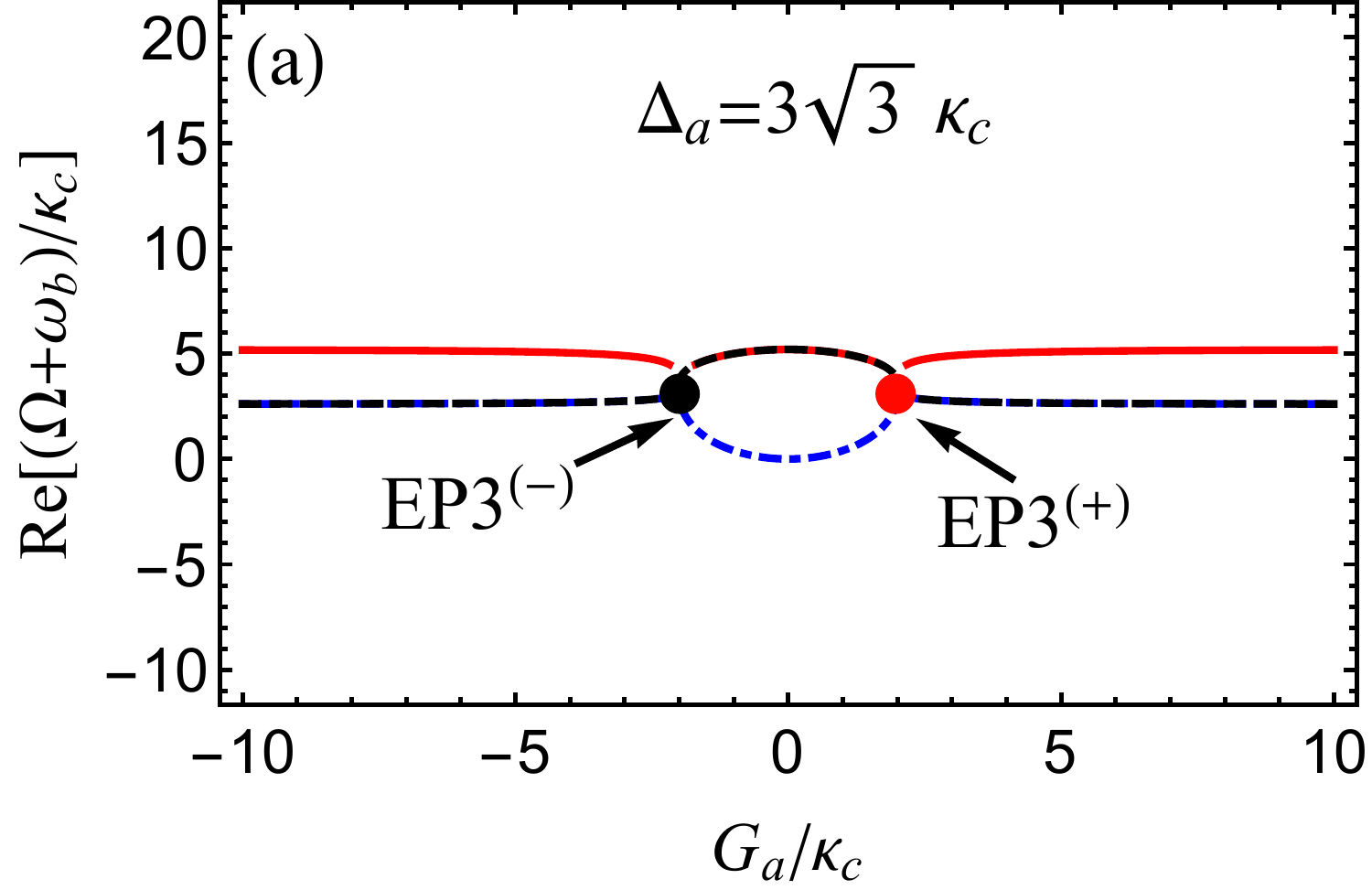}
	\includegraphics[scale=0.38]{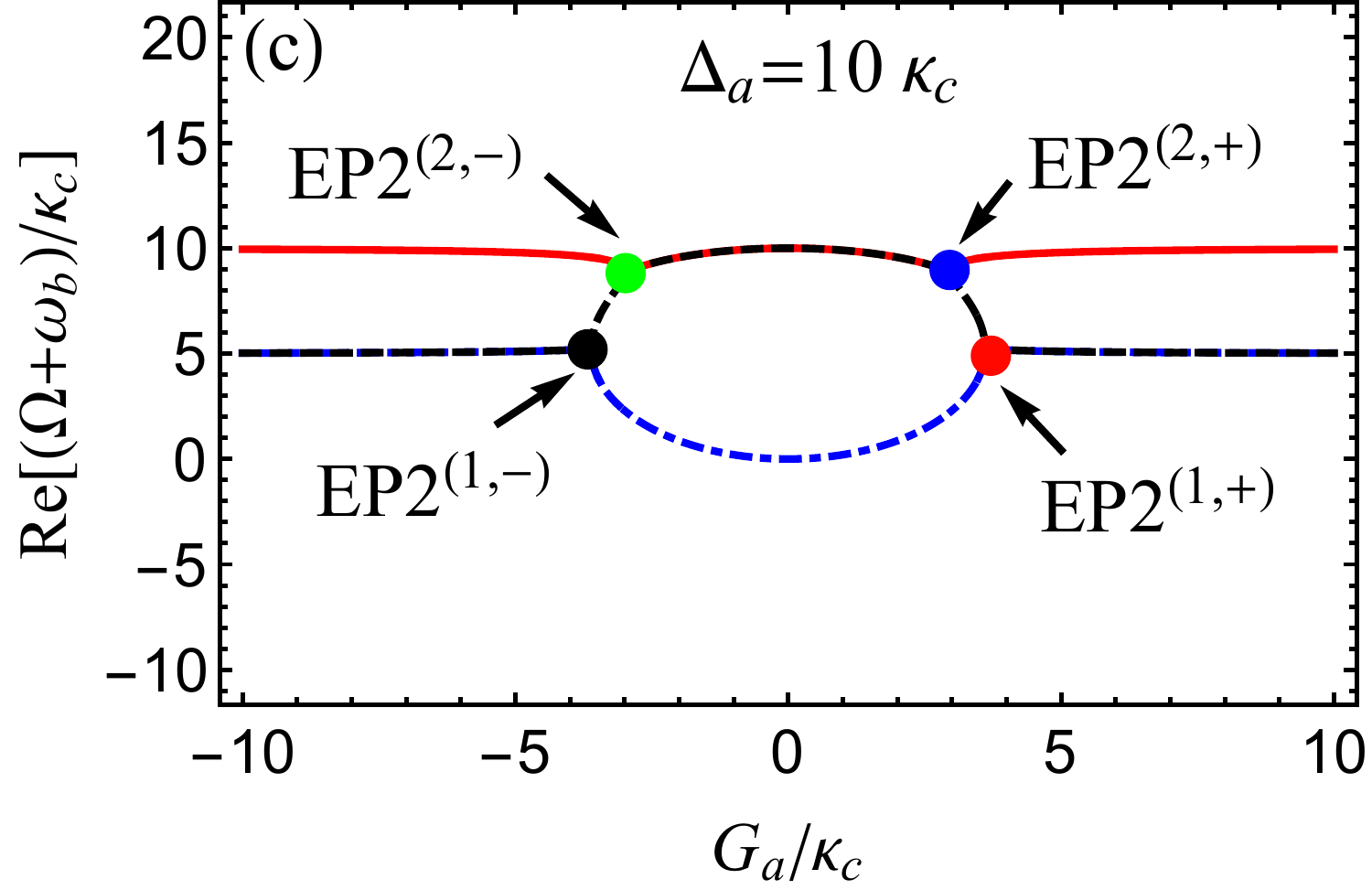}
	\includegraphics[scale=0.38]{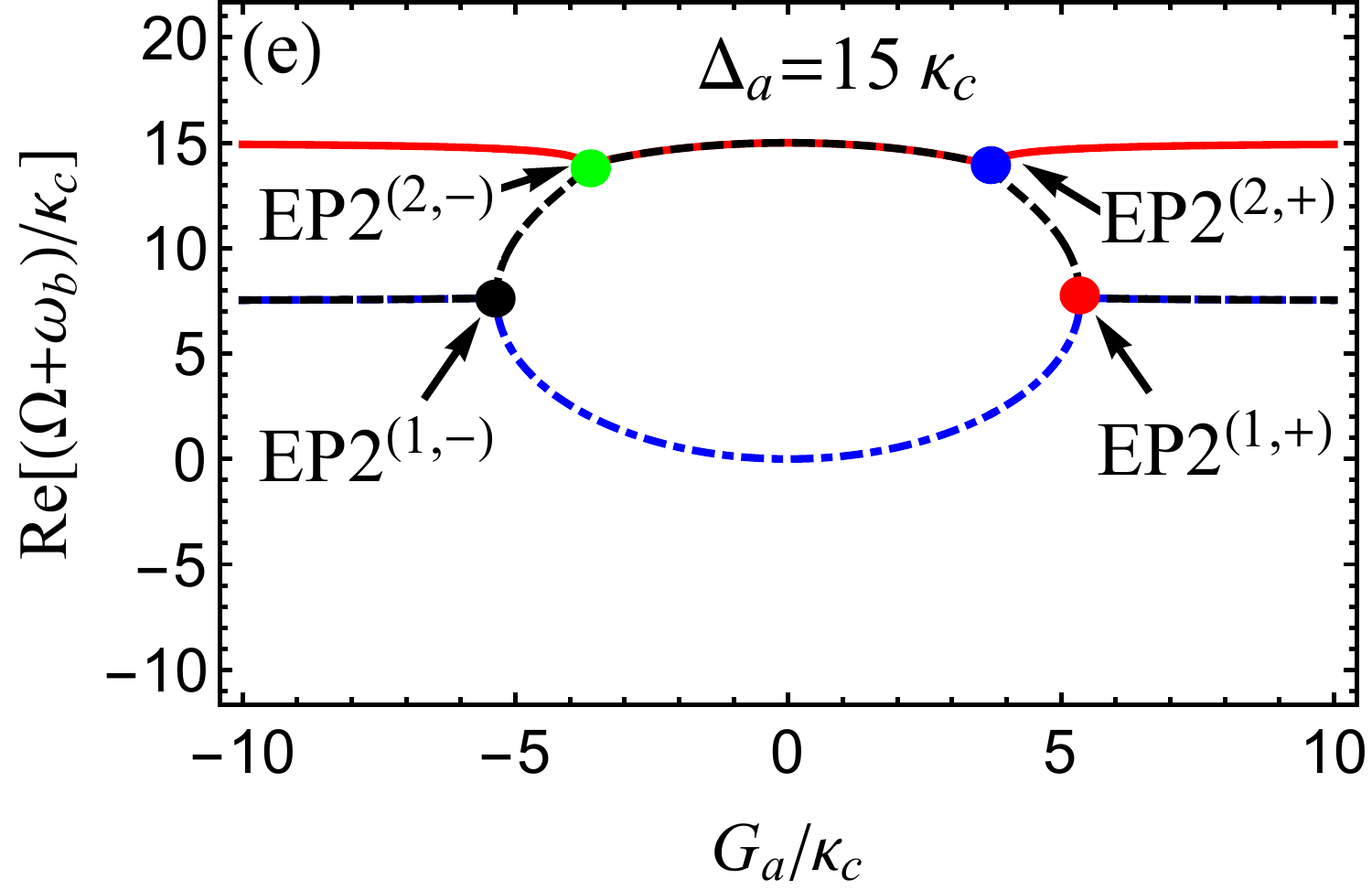}
	\includegraphics[scale=0.38]{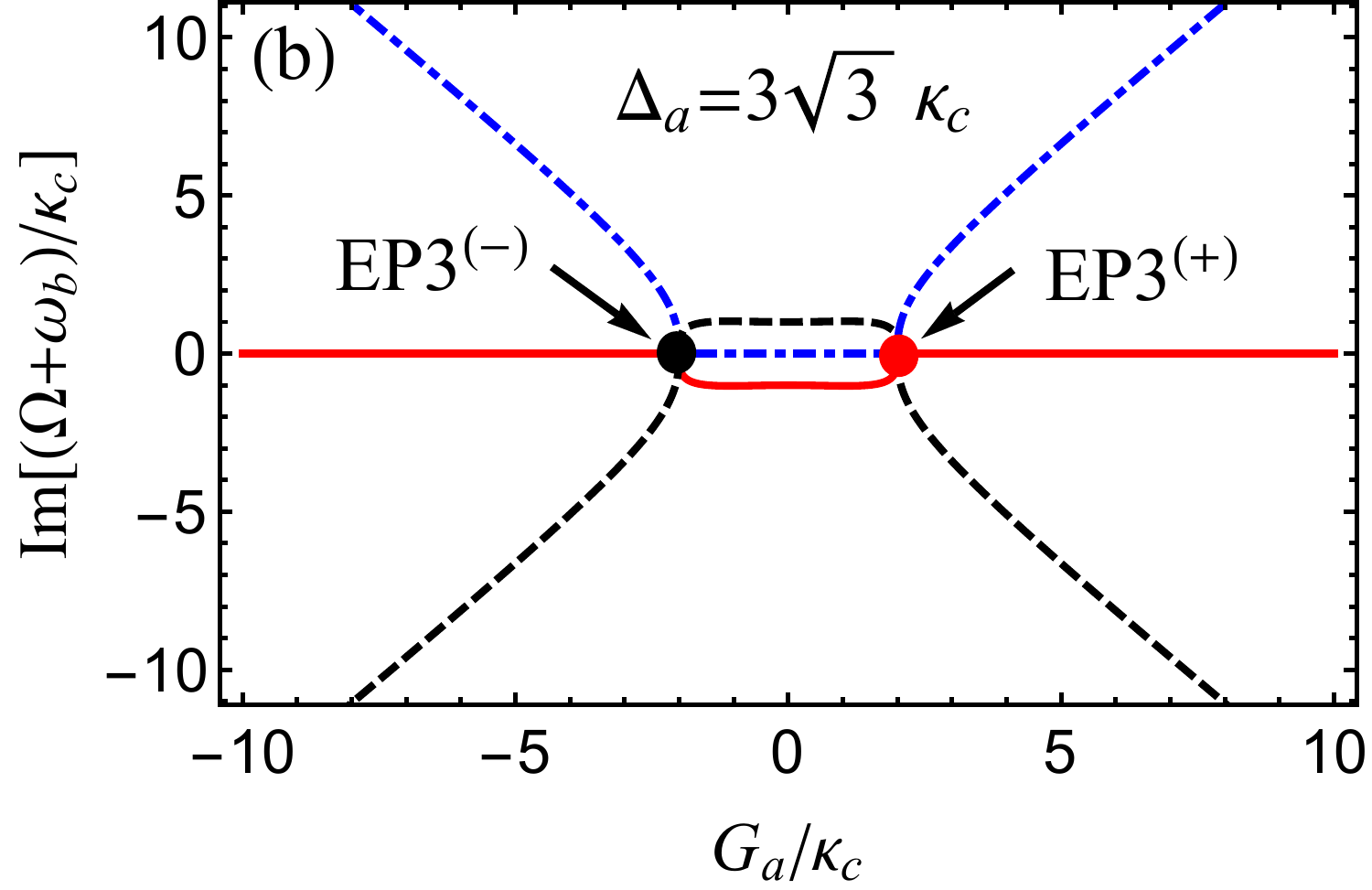}
	\includegraphics[scale=0.38]{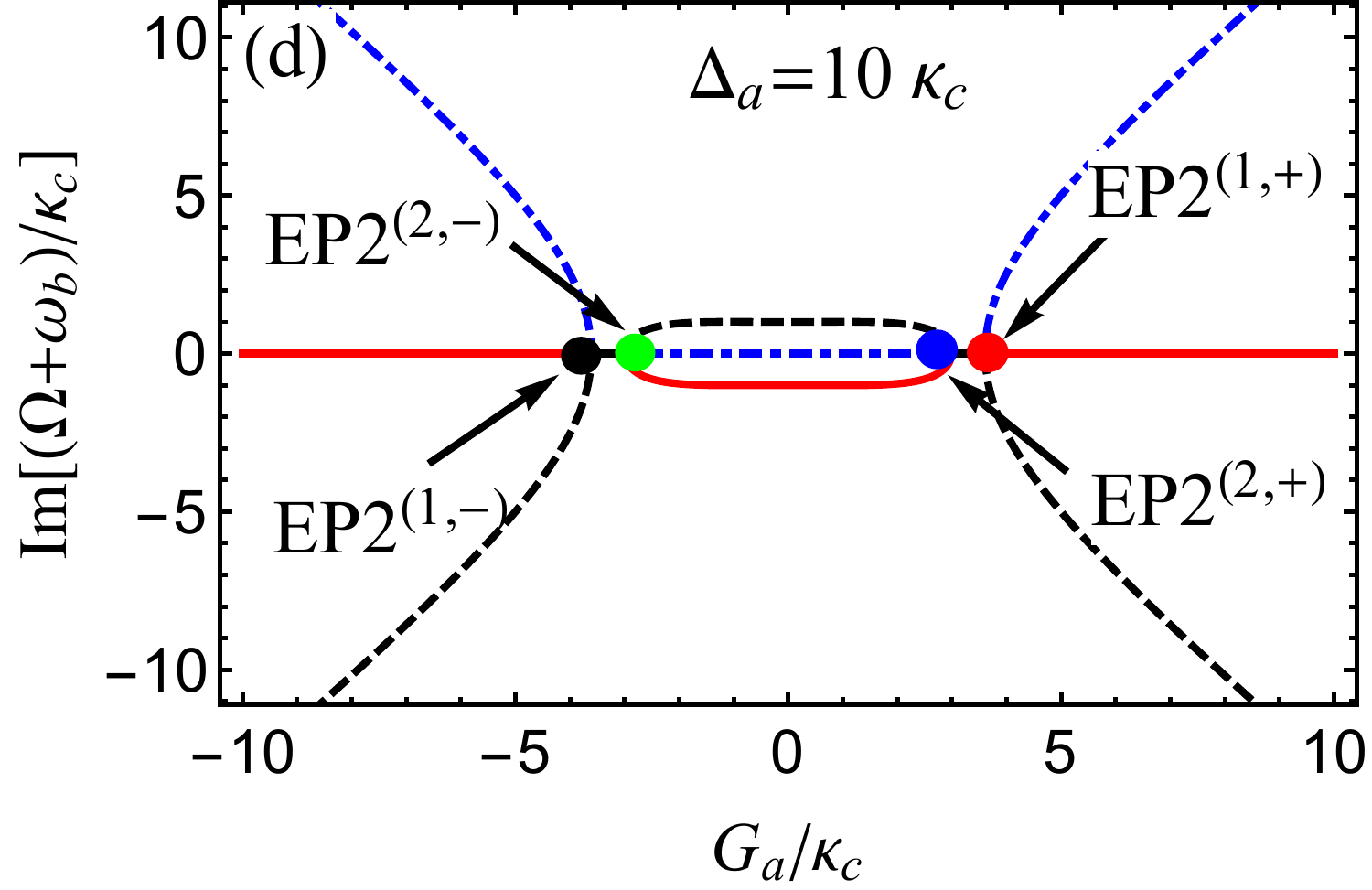}
	\includegraphics[scale=0.38]{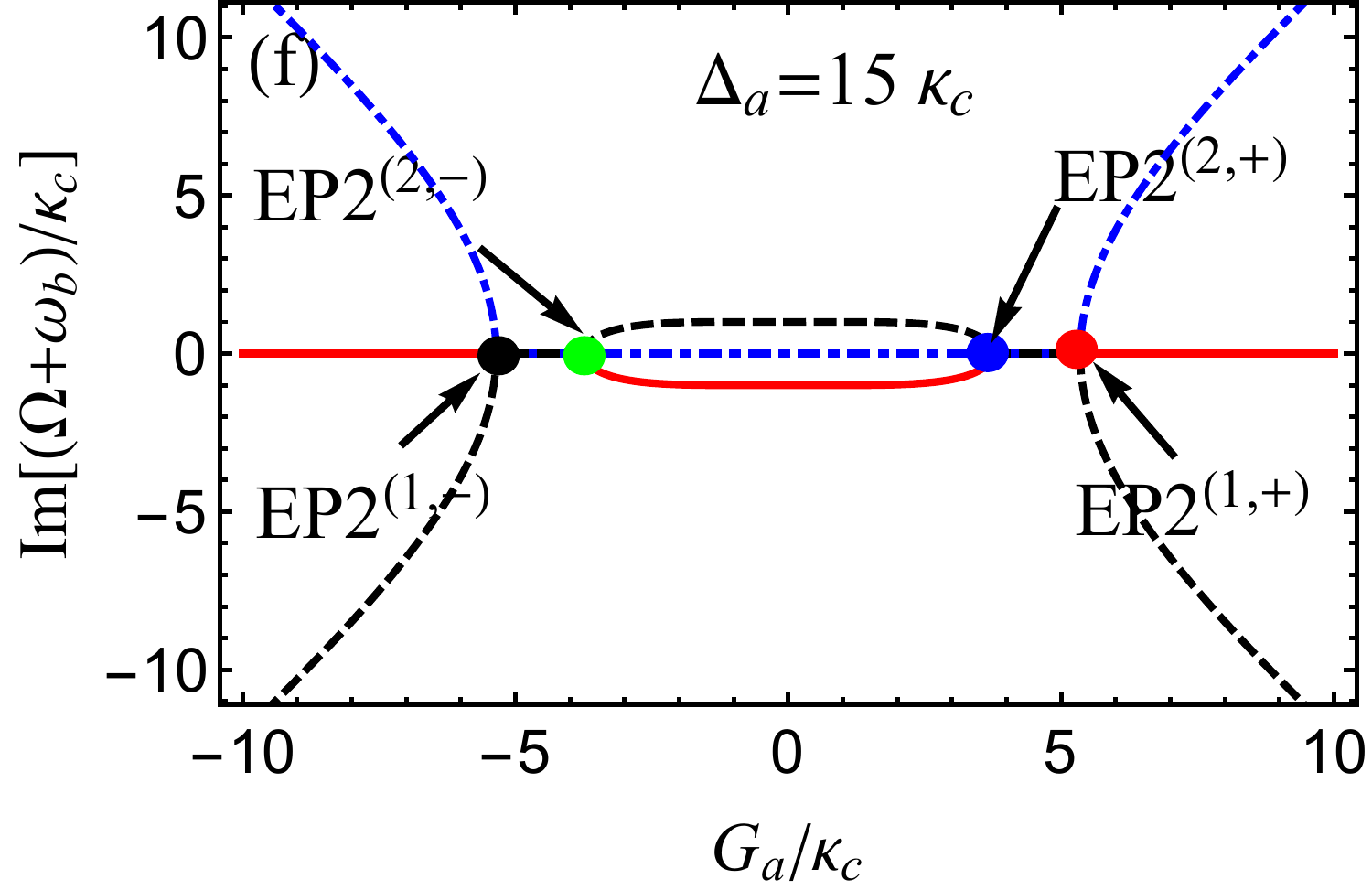}
	\caption{ The real and imaginary parts of the eigenvalue ($x=\Omega+\omega_b$) as a function of the normalized parameter $G_a/\kappa_c$ with (a) $\Delta_a=3\sqrt{3}\kappa_c$, (b) $\Delta_a=10\kappa_c$, and (c) $\Delta_a=15\kappa_c$. Here $\eta=-1$ and $\lambda=1$.}\label{fig3}
\end{figure*}

\begin{figure*}
	\includegraphics[scale=0.38]{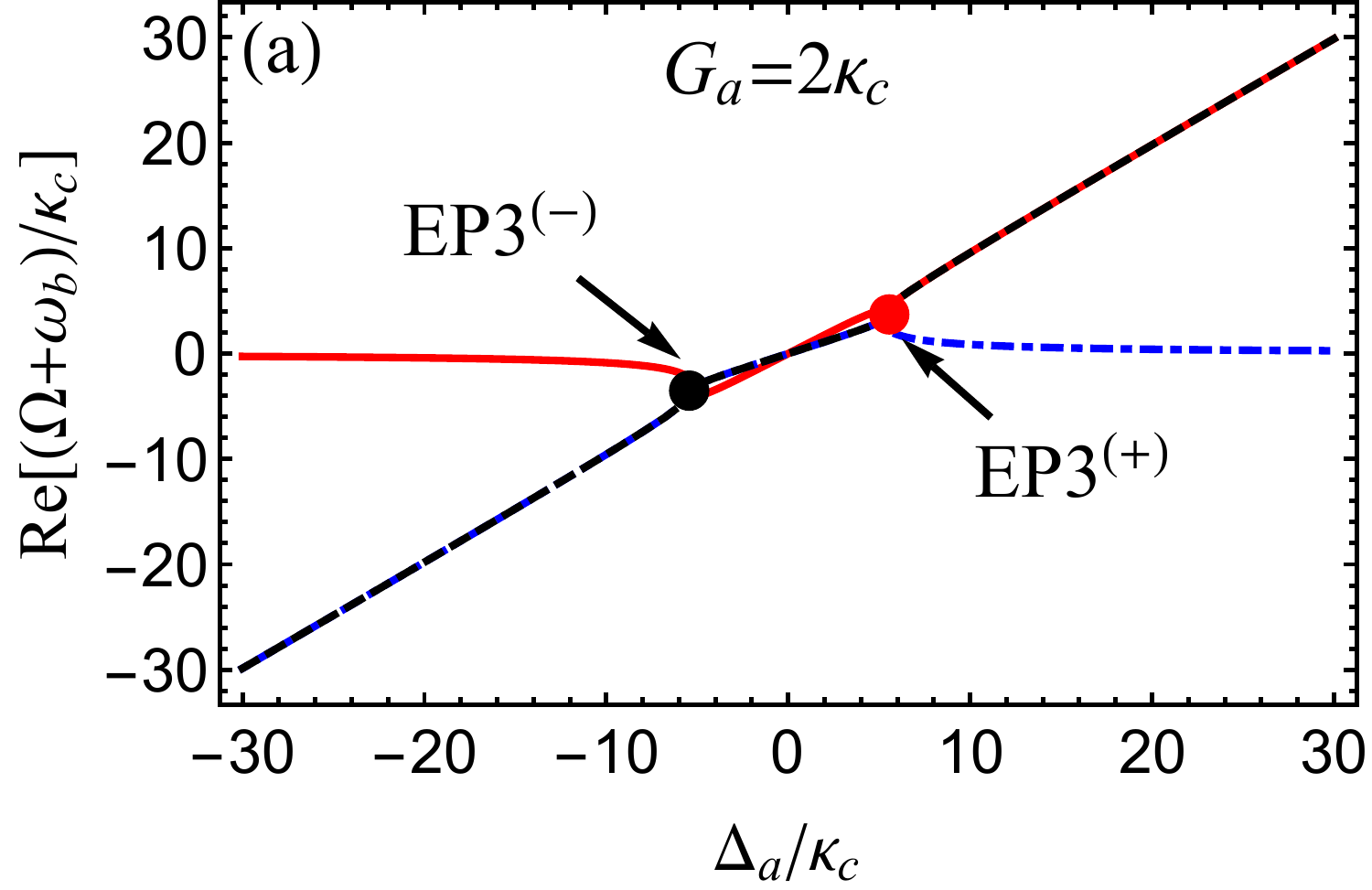}
	\includegraphics[scale=0.38]{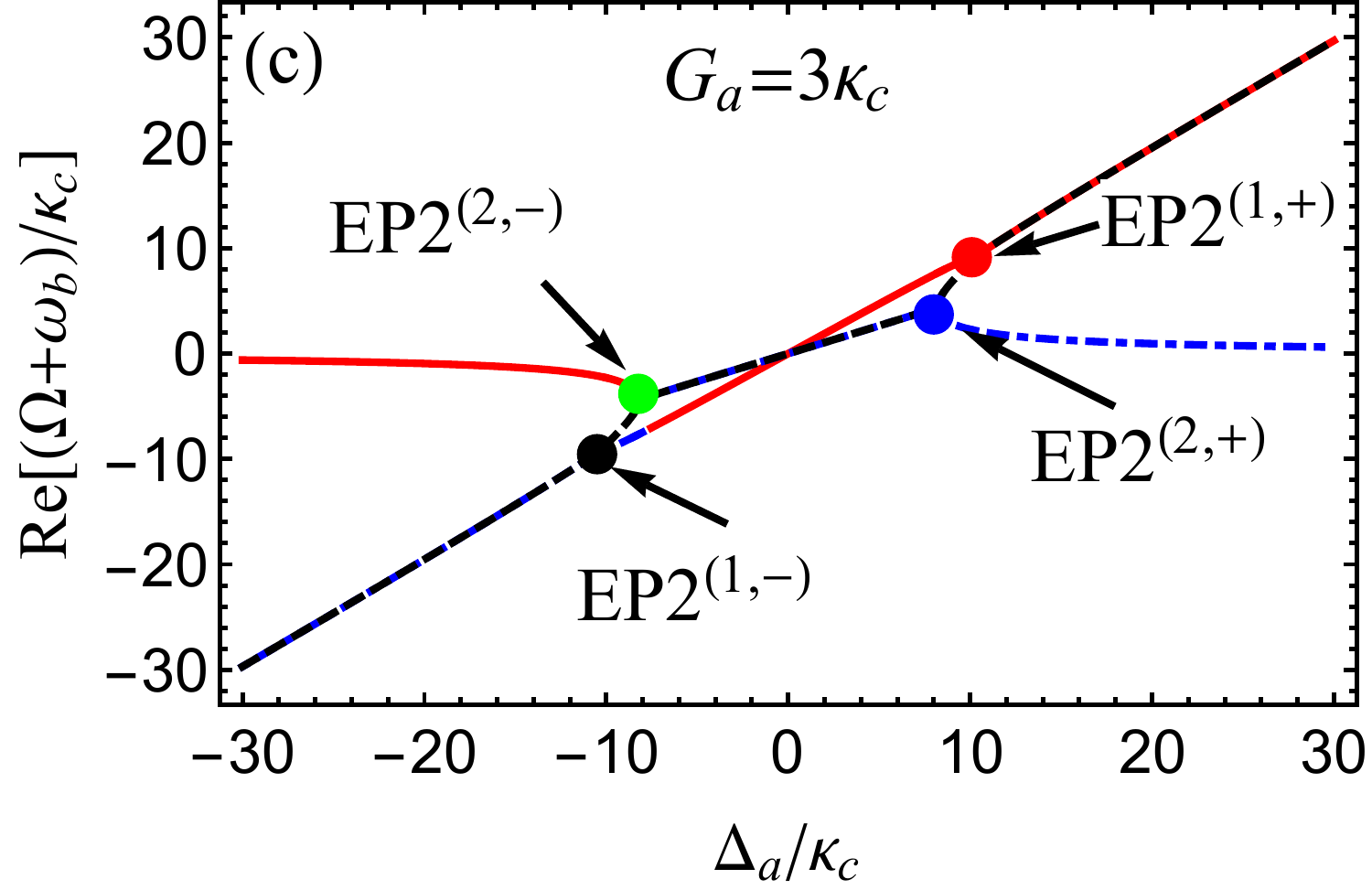}
	\includegraphics[scale=0.38]{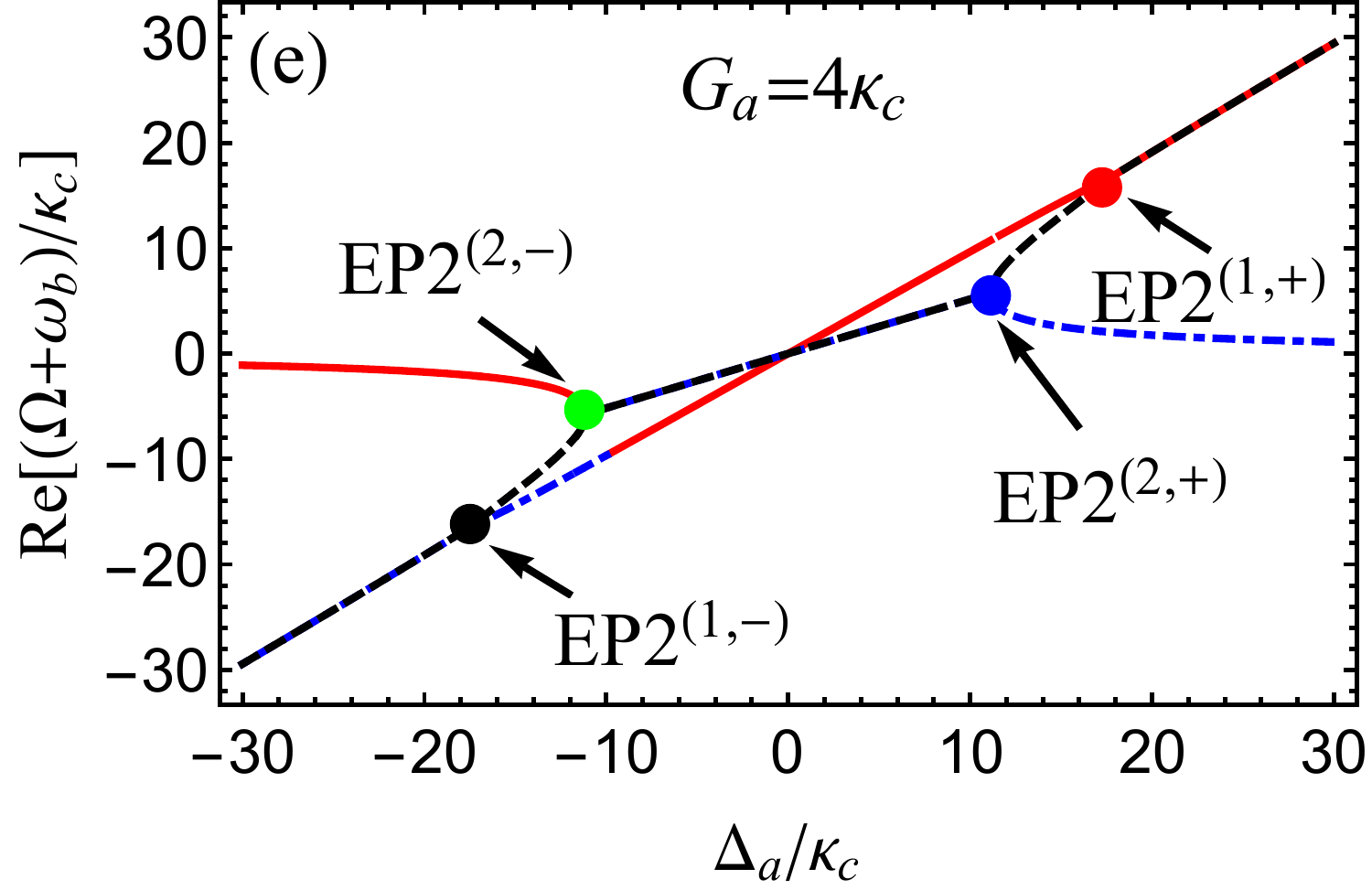}
	\includegraphics[scale=0.38]{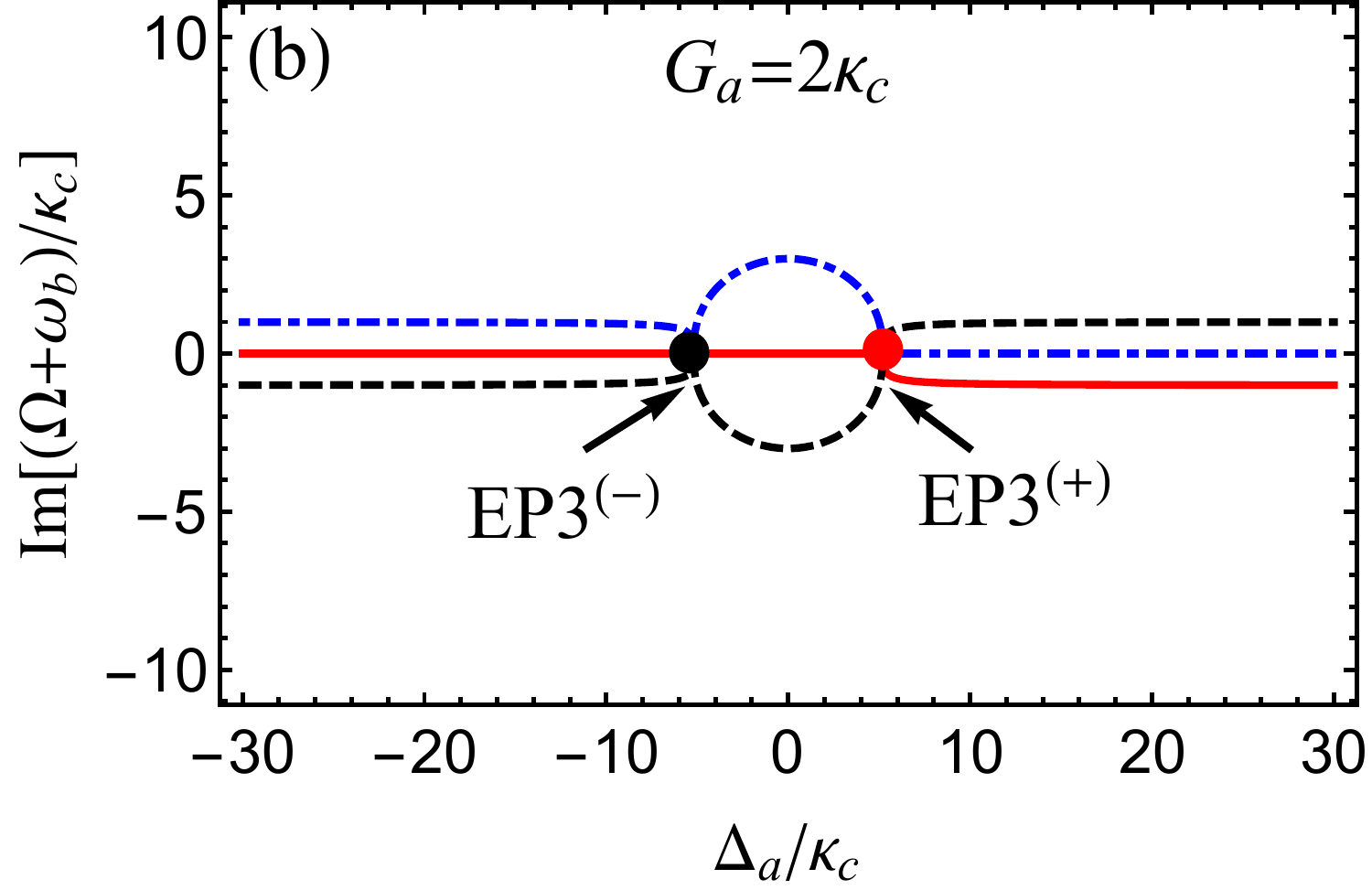}
	\includegraphics[scale=0.38]{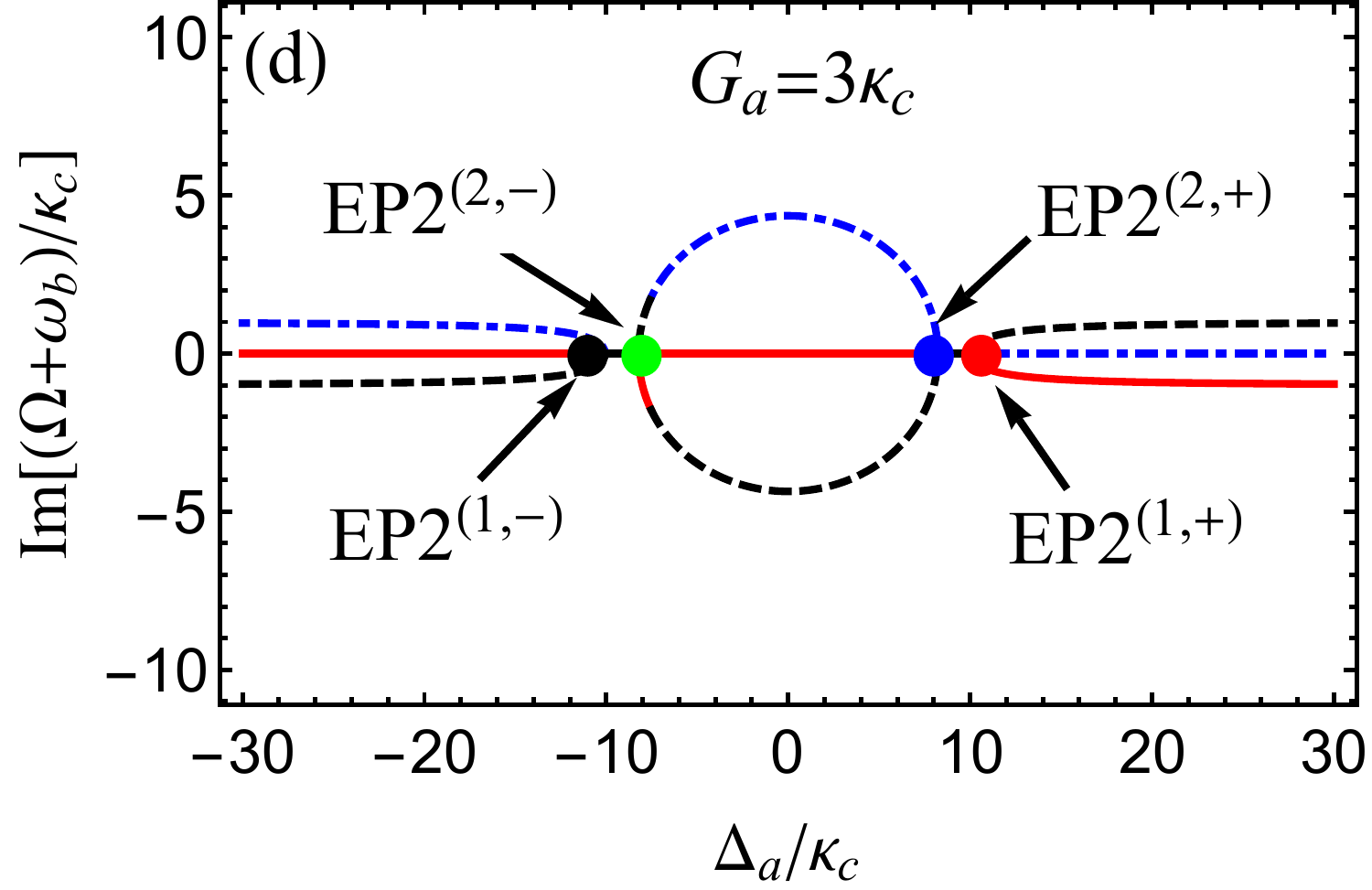}
	\includegraphics[scale=0.38]{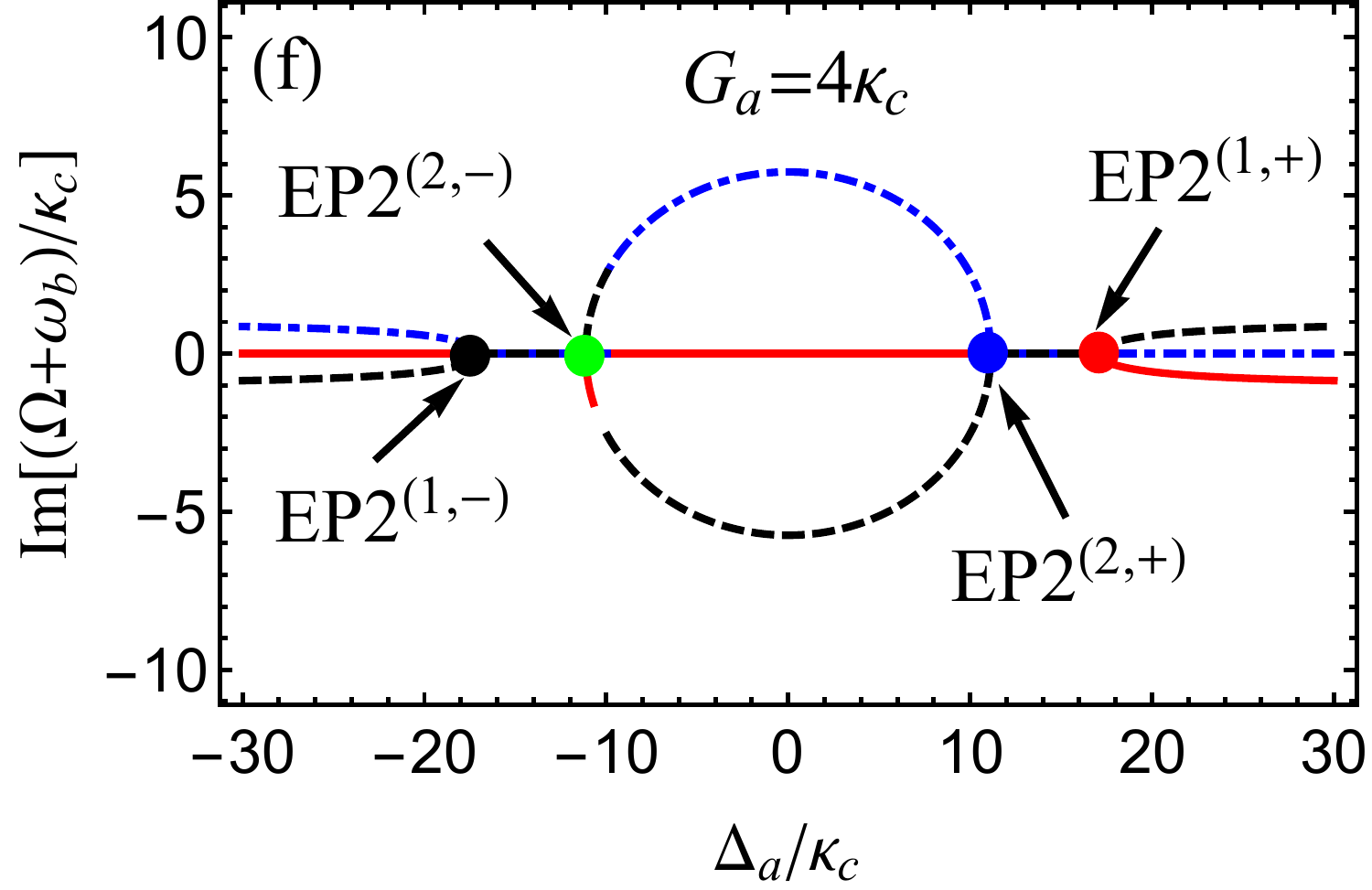}
	\caption{The real and imaginary parts of the eigenvalue ($x=\Omega+\omega_b$) as a function of the normalized parameter $\Delta_a/\kappa_c$ with (a) $G_a=2\kappa_c$, (b) $G_a=3\kappa_c$, and (c) $G_a=4\kappa_c$. The other parameters are the same as in Fig.~\ref{fig3}.}\label{fig4}
\end{figure*}
\subsection{$\eta\neq-1$}
For the more realistic case, $\eta\neq-1$ is further considered. This indicates that two optomechanical cavities are gain-loss unbalanced. According to Eq.~(\ref{eq36}), we can discuss the case of $\eta\neq-1$ in two scenarios, i.e., (i) $-2<\eta<-1$ (or $\gamma_b>0$); (ii) $-1<\eta<-\frac{1}{2}$ (or $\gamma_b<0$). The first scenario indicates that the loss MR and $\lambda>1$ (i.e., $|G_c|>|G_a|$) are needed to predict EP3 in our proposed COM system. On the contrary, the gain MR and $\lambda<1$ (i.e., $|G_c|<|G_a|$) is required in the second scenario. As examples, we take $\eta=-1.1$  and $\eta=-0.8$ [see the blue and green dots in Fig.~\ref{fig1_1}]. Then we plot the phase diagram of the discriminant with $\eta=-1.1$ [see Fig.~\ref{fig2}(b)] and $-0.8$ [see Fig.~\ref{fig2}(c)] vs the normalized parameters $G_a/\kappa_c$ and $G_c/\kappa_c$, where $D=0$, $A=0$, and $B=0$ are shown by the red, black and green curves, respectively. The purple (yellow) region means $D<0~(D>0)$. Obviously, four EP3s [see the red dots] produced by three curves, at which $D=A=B=0$, can be predicted in both Figs.~\ref{fig2}(b) and \ref{fig2}(c) by tuning $G_a$ and $G_c$. This can be realized because both $G_a$ and $G_c$ are tunable coupling strengths via tuning the Rabi frequencies of the two laser fields. When we deviate $G_a$ ($G_b$) from $G_{\rm a, EP3}$ ($G_{\rm b, EP3}$) at EP3, EP2 emerges [see the red curve only in Figs.~\ref{fig2}(b) and \ref{fig2}(c)]. When one parameter is fixed in Figs.~\ref{fig2}(a-c), we can easily find that only two EP3s or four EP2s can be observed by varying the other parameter, which is different from the case considered in the red-detuned COM system.

\section{EP3 and EP2 in the blue-sideband three-mode optomechanical system}

\begin{figure*}
	\includegraphics[scale=0.36]{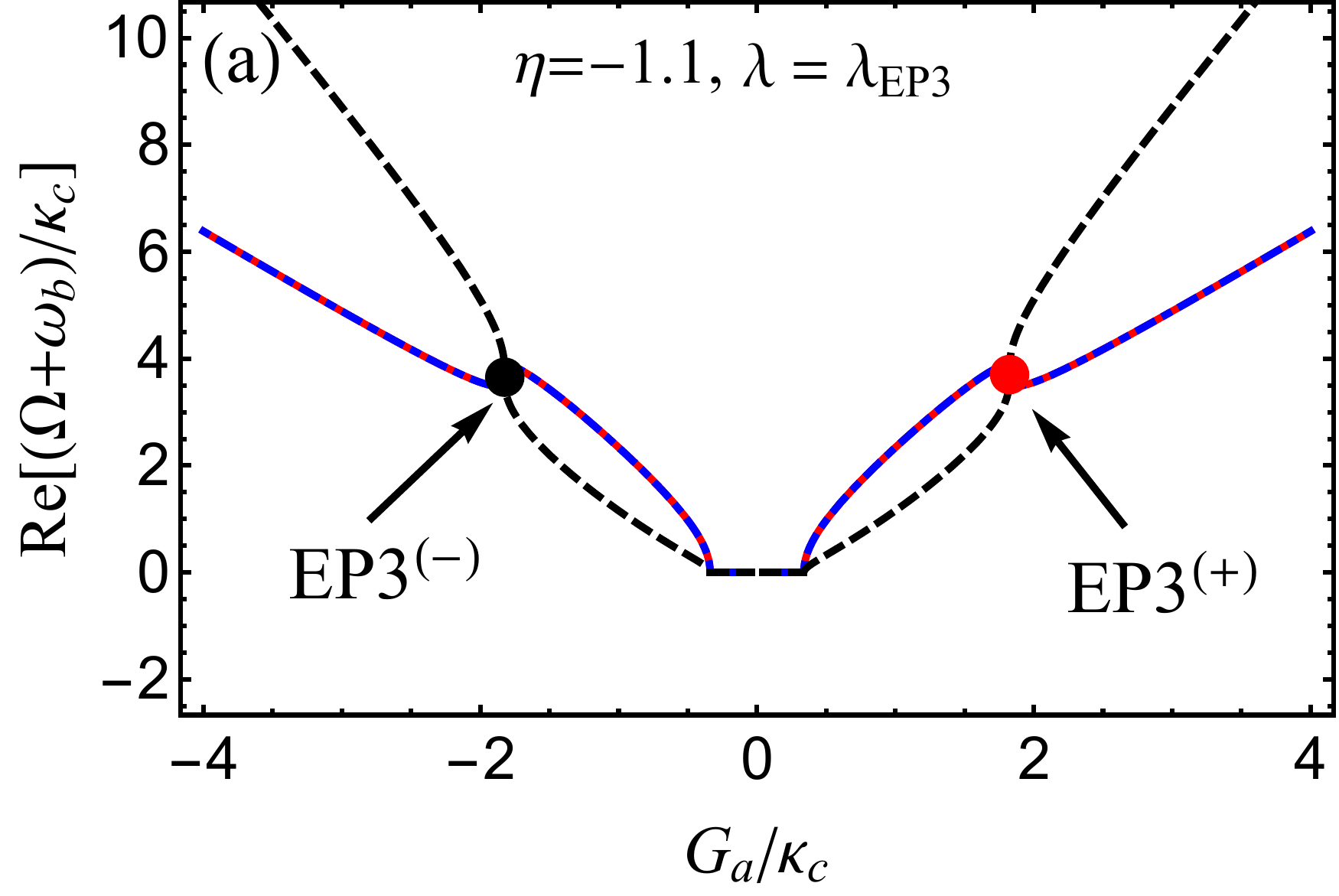}
	\includegraphics[scale=0.385]{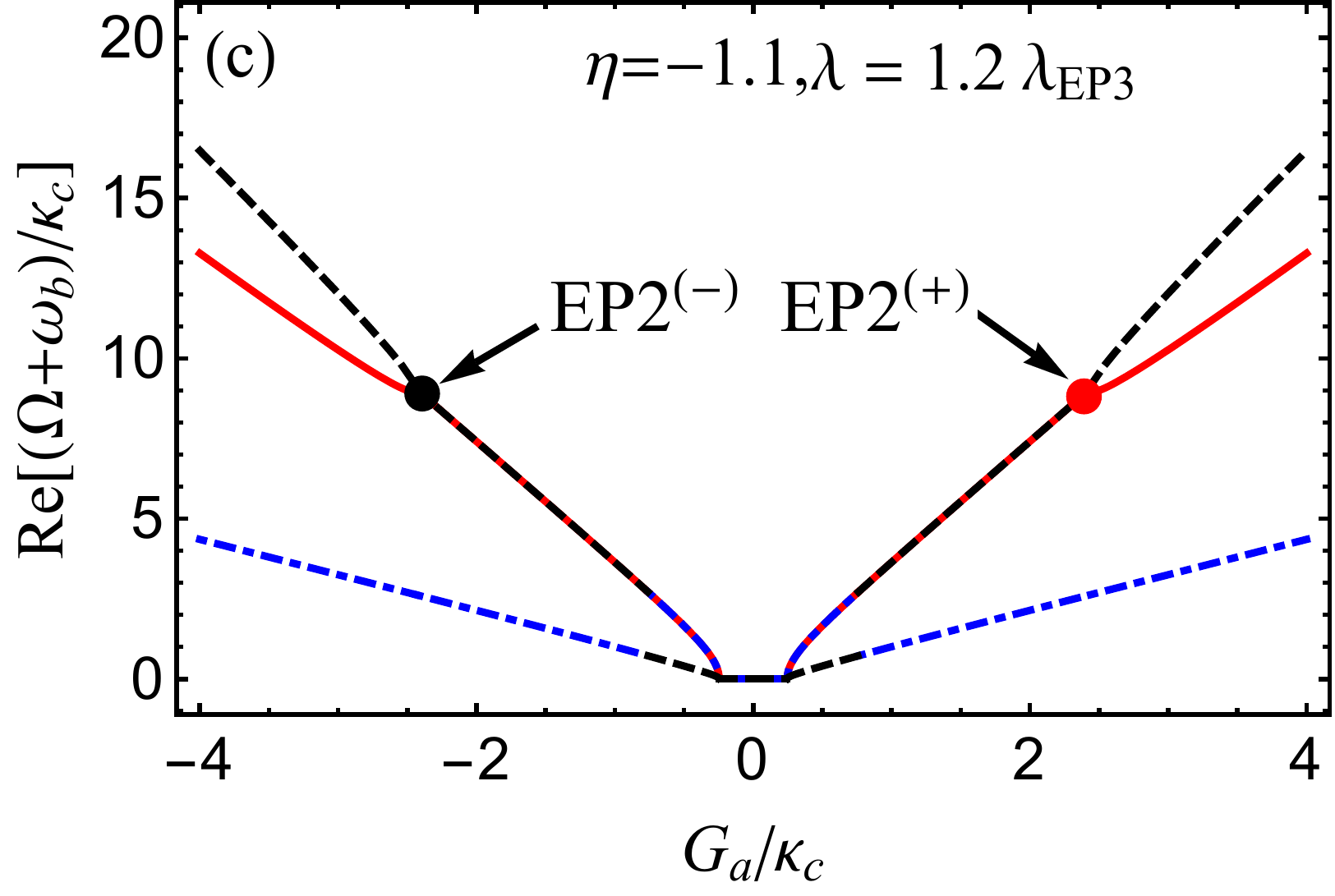}\\
	\includegraphics[scale=0.36]{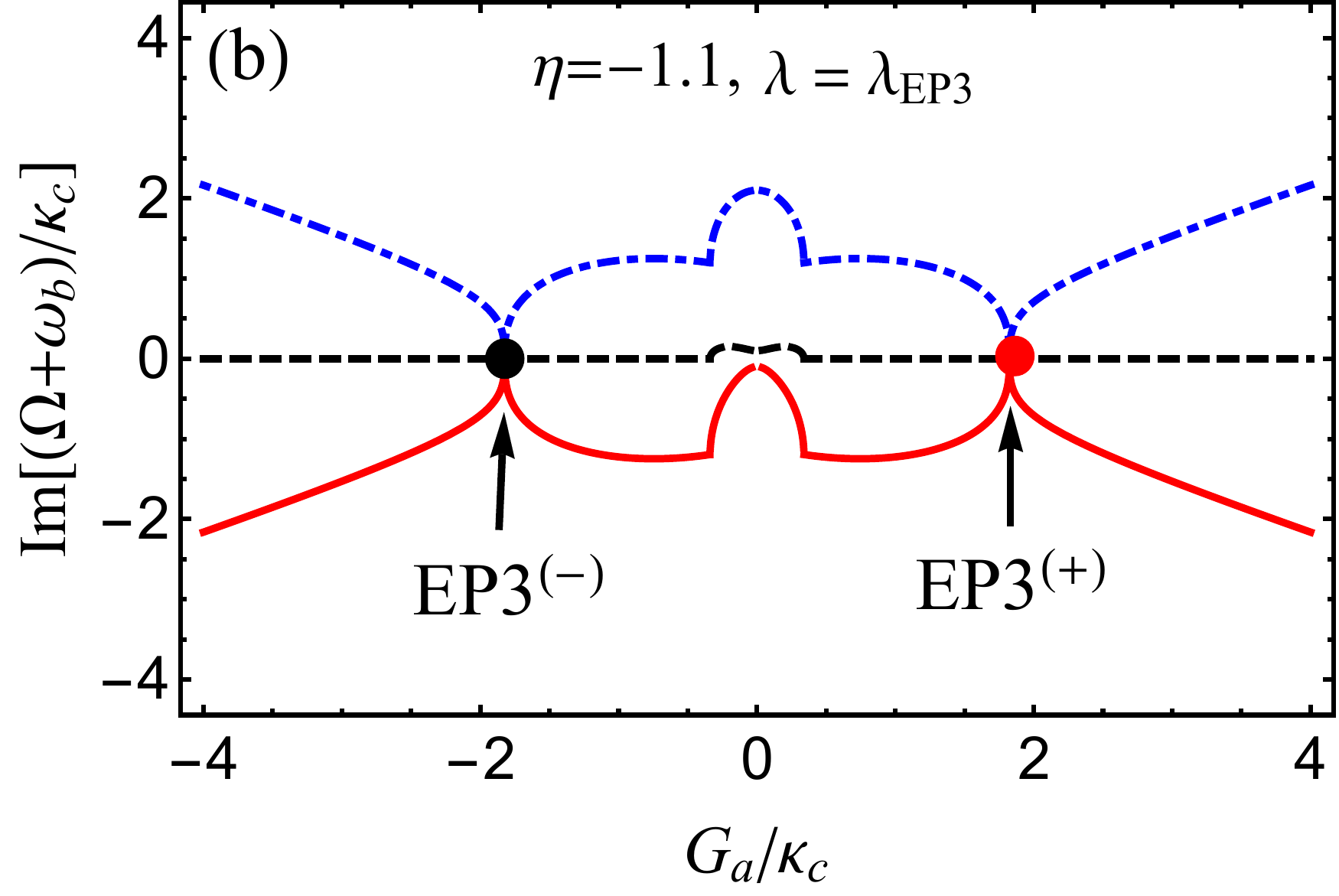}
	\includegraphics[scale=0.385]{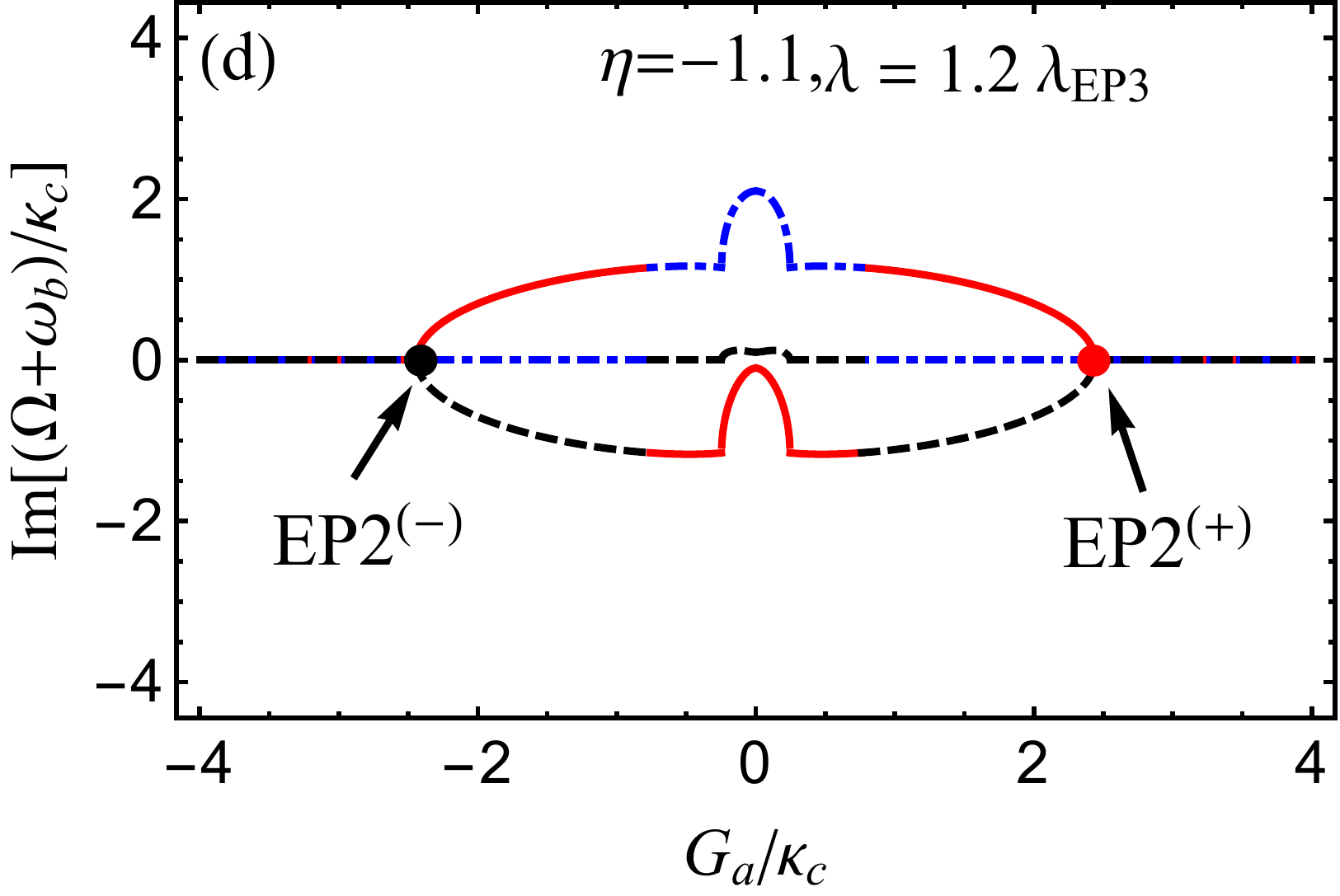}
	\caption{The real and imaginary parts of the eigenvalue ($x=\Omega+\omega_b$) with $\eta=-1.1$ as a function of the normalized parameter $G_a/\kappa_c$ with different $\lambda$. In (a) and (b), $\lambda=\lambda_{\rm EP3}$. In (c) and (d), $\lambda=1.2\lambda_{\rm EP3}$. }\label{fig5}
\end{figure*}
 
\subsection{$\eta=-1$}

In Fig.~\ref{fig2}(a), we have predicted that EP3 and EP2 can be observed in our considered system for the case of $\eta=-1$. {For clarity}, we below study the behavior of three eigenvalues of the Hamiltonian $H_{\rm eff}$ given by Eq.~(\ref{eq16}) with different frequency detunings $\Delta_a$ and coupling strengths $G_a$.

In Fig.~\ref{fig3}, we plot the real and imaginary parts of the eigenvalue ($x=\Omega+\omega_b$) as a function of the normalized parameter $G_a/\kappa_c$ with $\Delta_a=3\sqrt{3}\kappa_c,~10\kappa_c,~15\kappa_c$. For simplicity, we assume that the red, blue and black curves respectively denote three eigenvalues $x_1,~x_2$ and $x_3$ for the Hamiltonian $H_{\rm eff}$ hereafter. For $\Delta_a=3\sqrt{3}\kappa_c$ [see Figs.~\ref{fig3}(a) and \ref{fig3}(b)], $x_1$ is real and the other two eigenvalues ($x_2$ and $x_3$) are a complex-conjugate pair in the region of $G_a>G_{\rm a, EP3}^{(+)}=2\kappa_c$ or $G_a<G_{\rm a, EP3}^{(-)}=-2\kappa_c$. But when  $G_{\rm a, EP3}^{(-)}<G_a<G_{\rm a, EP3}^{(+)}$, $x_2$ becomes real, and $x_1$, $x_3$ become a complex-conjugate pair. At the points $G_a=G_{\rm a, EP3}^{(\pm)}$ [see the red and black dots in Figs.~\ref{fig3}(a) and \ref{fig3}(b)], three eigenvalues coalesce to one value, i.e., $\Omega_{\rm EP3}=3.39\kappa_c-\omega_b$, corresponding to two degenerate EP3s. It is not difficult to verify that $D=0,~A=0$ and $B=0$ at two EP3s. Then we increase $\Delta_a$ to $\Delta_a=10\kappa_c$ [see Figs.~\ref{fig3}(c) and \ref{fig3}(d)] for deviation from EP3s, that is, $D=0$ but $A\neq0$ and $B\neq0$. For $G_a<G_{\rm a, EP3}^{(1,-)}=-3.6\kappa_c$ or $G_a>G_{\rm a, EP3}^{(1,+)}=3.6\kappa_c$ , $x_1$ is real, $x_2$ and $x_3$ are a complex-conjugate pair, At $G_a=G_{\rm a, EP3}^{(1,\pm)}$ [see the red and black dots in Figs.~\ref{fig3}(c) and \ref{fig3}(d)], $x_2$ and $x_3$ coalesce to $\Omega_{\rm EP2}^{(1)}=5.22\kappa_c-\omega_b$, corresponding to two degenerate EP2s. By increasing $G_a$ to $G_{\rm a, EP3}^{(1,-)}<G_a<G_{\rm a, EP3}^{(2,-)}=-2.99\kappa_c$ or $2.99\kappa_c=G_{\rm a, EP3}^{(2,+)}<G_a<G_{\rm a, EP3}^{(1,+)}=3.6\kappa_c$, the real parts of $x_2$ and $x_3$ bifurcate into two values. At $G_a=G_{\rm a, EP3}^{(2,\pm)}$ [see the blue and green dots in Figs.~\ref{fig3}(c) and \ref{fig3}(d)], $x_1$ and $x_3$ coalesce to the value $\Omega_{\rm EP2}^{(2)}=2.28\kappa_c-\omega_b$, corresponding another two degenerate EP2s. When $G_{\rm a, EP3}^{(2,-)}<G_a<G_{\rm a, EP3}^{(2,+)}$, $x_2$ is real and the other two eigenvalues $x_1$ and $x_3$ are complex conjugates. We also find the separation between arbitrary two EP2s can be increased using larger $\Delta_a$ such as $\Delta_a=15\kappa_c$~[see Figs.~\ref{fig3}(c-f)], which indicates that larger $\Delta_a$ is benifit to distinguishably observe multiple EP2s. 

In Fig.~\ref{fig4}, we also plot the real and imaginary parts of the eigenvalue ($x=\Omega+\omega_b$) vs the normalized parameter $\Delta_a/\kappa_c$ with different $G_a$. For $G_a=2\kappa_c$ [see Figs.~\ref{fig4}(a) and \ref{fig4}(b)], $x_1$ is real, and $x_2$, $x_3$ are a complex-conjugate pair when $\Delta_a<\Delta_{\rm a,EP3}^{(-)}=-5.2\kappa_c$. At $\Delta_{\rm a,EP3}^{(-)}=-5.2\kappa_c$ [see the black dot in Figs.~\ref{fig4}(a) and \ref{fig4}(b)], three eigenvalues coalesce to $\Omega_{\rm EP3}^{(1)}=3.55\kappa_c-\omega_b$, corresponding to EP3. When $\Delta_{\rm a,EP3}^{(-)}<\Delta_a<\Delta_{\rm a,EP3}^{(+)}=5.2\kappa_c$, $x_2$ and $x_3$ become a complex-conjugate pair again, and $x_1$ is real. But when $\Delta_a>\Delta_{\rm a,EP3}^{(+)}$, $x_2$ becomes real, $x_1$ and $x_3$ are a complex-conjugate pair. At $\Delta_a=\Delta_{\rm a,EP3}^{(+)}$, three eigenvalues remerge into the same value $\Omega_{\rm EP3}^{(2)}\approx3.55\Omega_{\rm EP3}^{(1)}$. By increasing $G_a$ to $G_a=3\kappa_c$ [see the red dot in Figs.~\ref{fig4}(c) and \ref{fig4}(d)], two EP3s in  Figs.~\ref{fig4}(a) and \ref{fig4}(b) split into four EP2s. Specifically, $x_1$ is real, $x_2$ and $x_3$ are a complex-conjugate pair when $G_a<G_{\rm a, EP2}^{(1,-)}=-10.1\kappa_c$. At $G_a=G_{\rm a, EP2}^{(1,-)}=-10.1\kappa_c$ [see the black dot in Figs.~\ref{fig4}(c) and \ref{fig4}(d)], $x_2$ and $x_3$ coalesce to $\Omega_{\rm EP2}^{(1)}=-2.27\kappa_c-\omega_b$, corresponding to the first EP2. When $G_{\rm a, EP2}^{(1,-)}<G_a<G_{\rm a, EP2}^{(2,-)}=-8.23\kappa_c$, three eigenvalues are all real but have different values. At $G_a=G_{\rm a, EP2}^{(2,-)}$ [see the green dot in Figs.~\ref{fig4}(c) and \ref{fig4}(d)], $x_1$ and $x_3$ coalesce to $\Omega_{\rm EP2}^{(2)}=-4.4\kappa_c-\omega_b$, corresponding to the second EP2. For $G_{\rm a, EP2}^{(2,-)}<G_a<G_{\rm a, EP2}^{(2,+)}=8.23\kappa_c$, $x_1$ is real, $x_2$ and $x_3$ are a complex-conjugate pair. At $G_a=G_{\rm a, EP2}^{(2,+)}$ [see the blue dot in Figs.~\ref{fig4}(c) and \ref{fig4}(d)], $x_2$ and $x_3$ remerge into one value $\Omega_{\rm EP2}^{(3)}\approx\Omega_{\rm EP2}^{(2)}$, corresponding to the third EP2. By tuning $G_a$ to $G_a=G_{\rm a, EP2}^{(1,+)}=10.1\kappa_c$ [see the red dot in Figs.~\ref{fig4}(c) and \ref{fig4}(d)], two different real eigenvalues (i.e., $x_1$ and $x_3$) in $G_{\rm a, EP2}^{(2,+)}<G_a<G_{\rm a, EP2}^{(1,+)}$ degenerate as $\Omega_{\rm EP2}^{(4)}=8.93\kappa_c-\omega_b$, corresping to the fourth EP2. When $G_a$ exceeds $G_{\rm a, EP2}^{(1,+)}$, $x_2$ becomes real, $x_2$ and $x_3$ are a complex conjugates. By considering a larger $G_a$ such as $G_a=15\kappa_c$ [see Figs.~\ref{fig4}(e) and \ref{fig4}(f)], we find EP2s can be distinguished more easily. This indicates that larger coupling strength can also be used to observe EP2s clearly, similar to the role of the above discussed frequency detuning $\Delta_a$.

\begin{figure*}
	\includegraphics[scale=0.36]{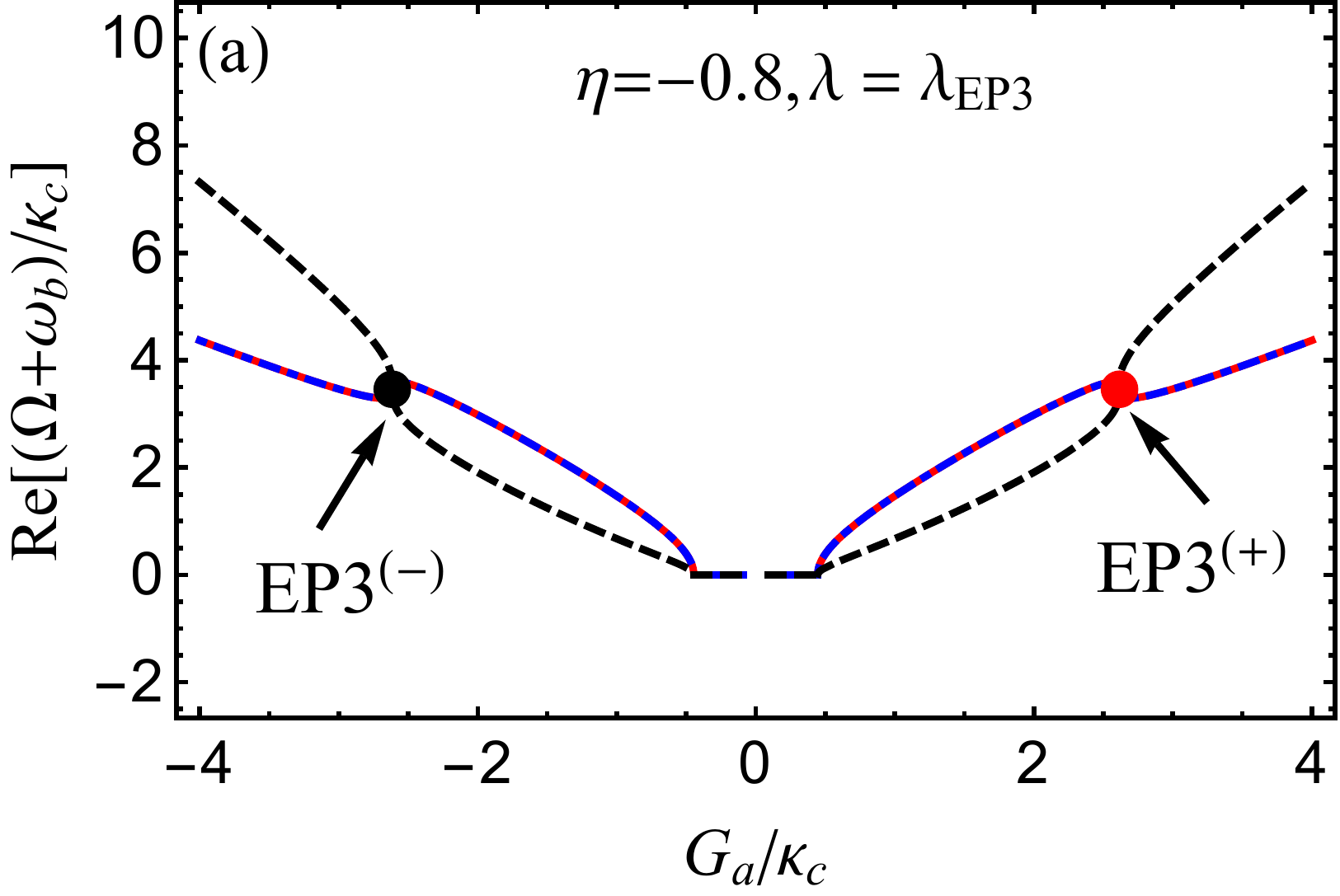}
	\includegraphics[scale=0.36]{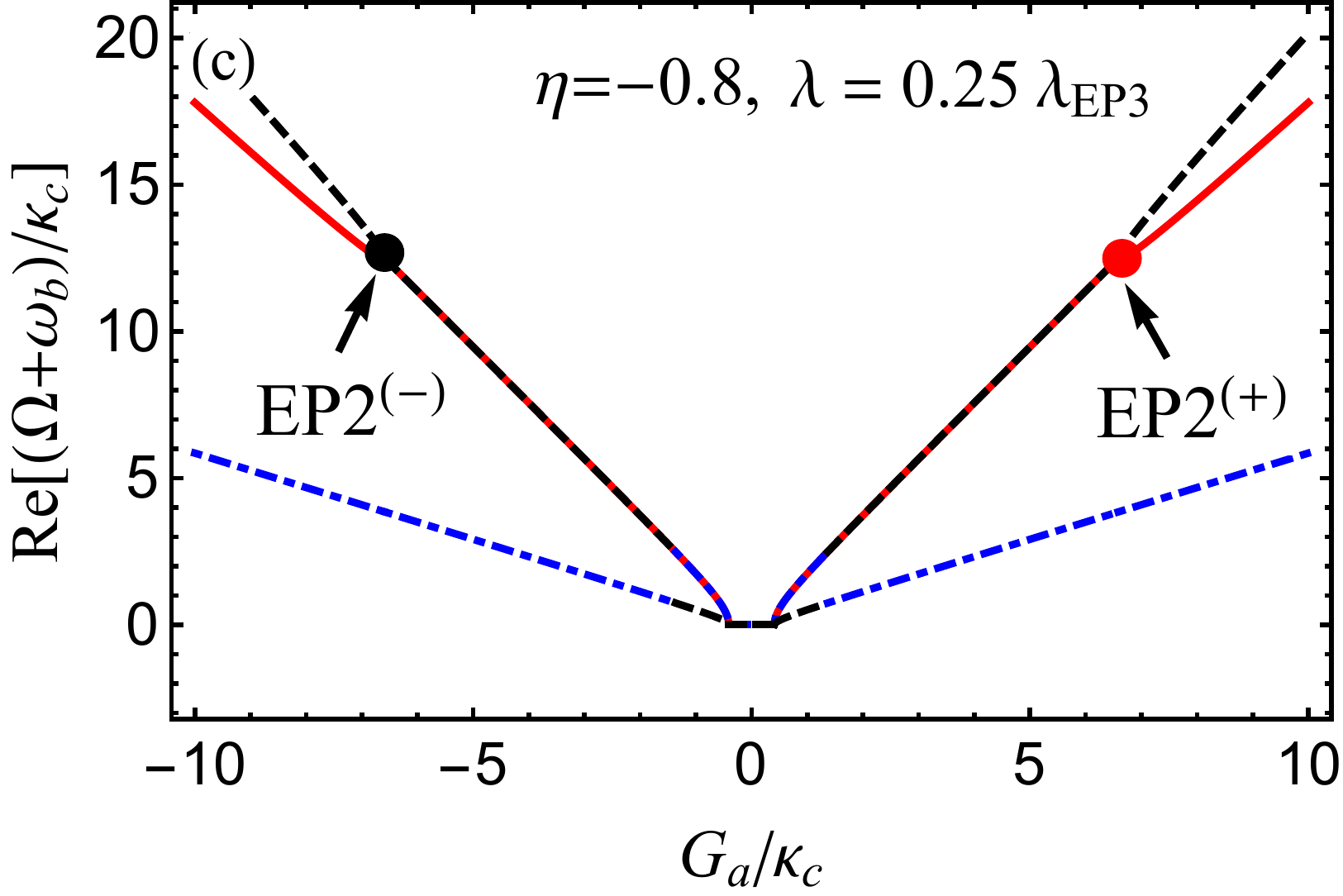}\\
	\includegraphics[scale=0.36]{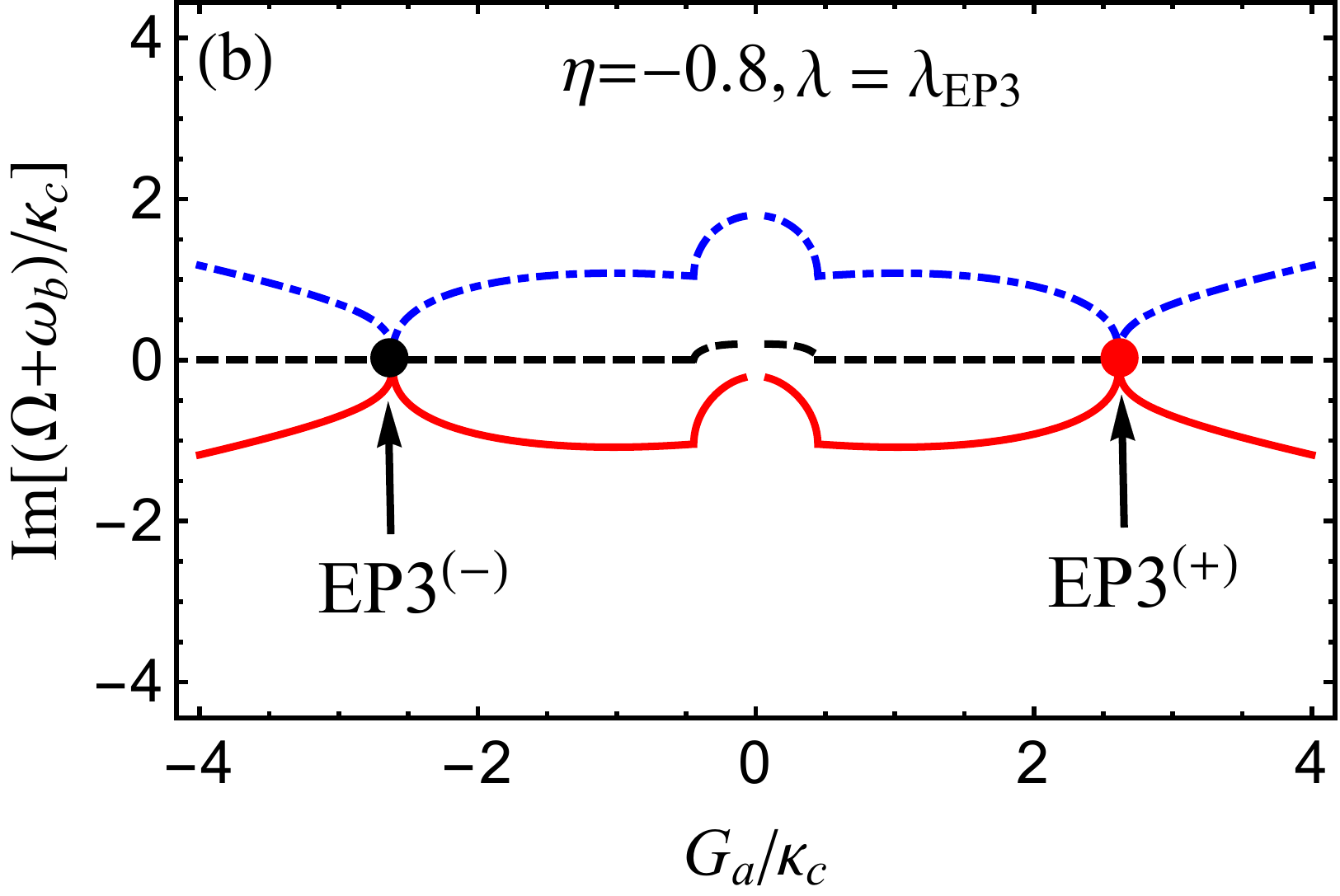}
	\includegraphics[scale=0.36]{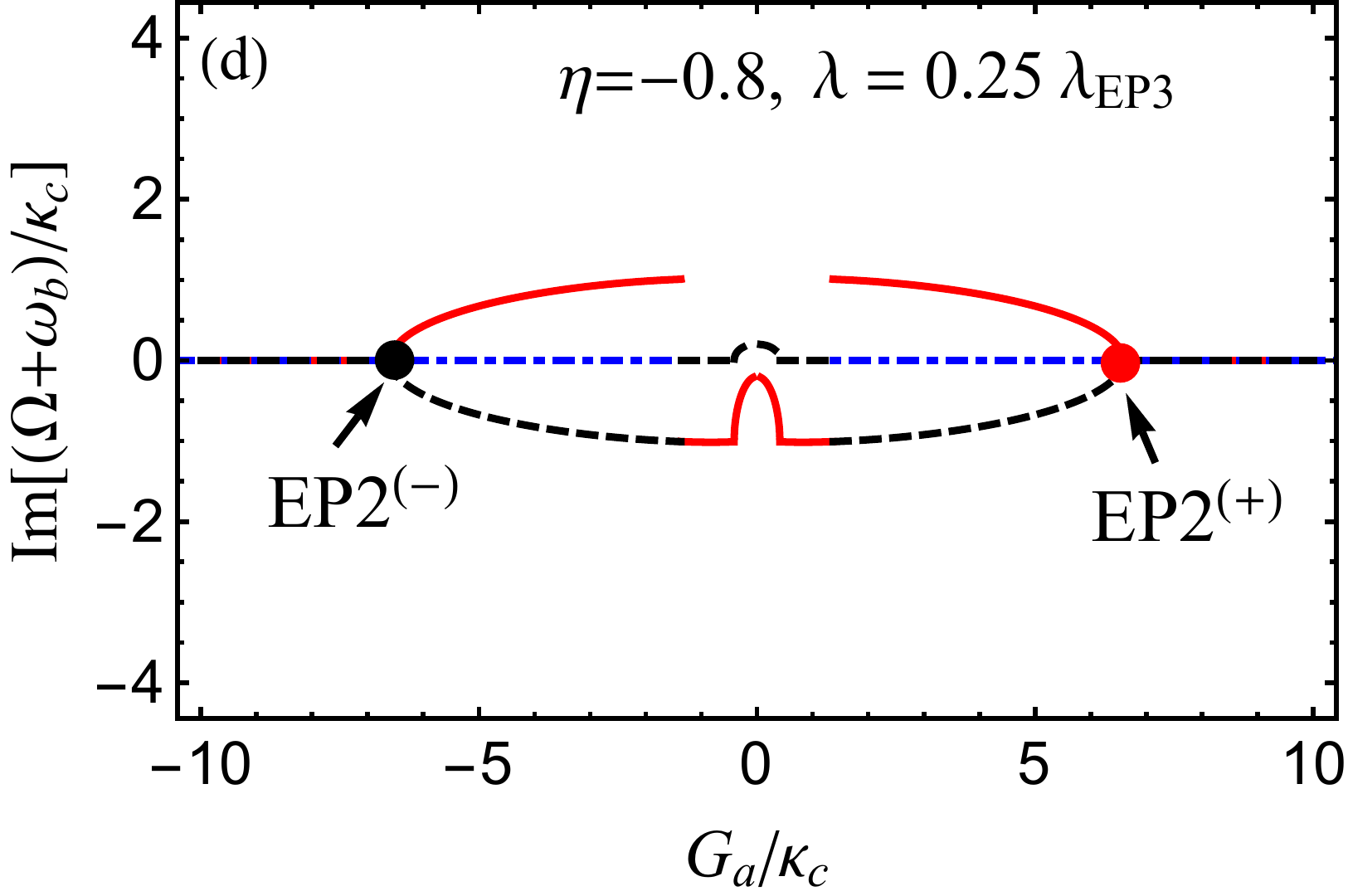}
	\caption{The real and imaginary parts of the eigenvalue ($x=\Omega+\omega_b$) with $\eta=-0.8$ as a function of the normalized parameter $G_a/\kappa_c$ with different $\lambda$. In (a) and (b), $\lambda=\lambda_{\rm EP3}$. In (c) and (d), $\lambda=0.25\lambda_{\rm EP3}$. }\label{fig6}
\end{figure*}
\begin{figure*}
	\includegraphics[scale=0.25]{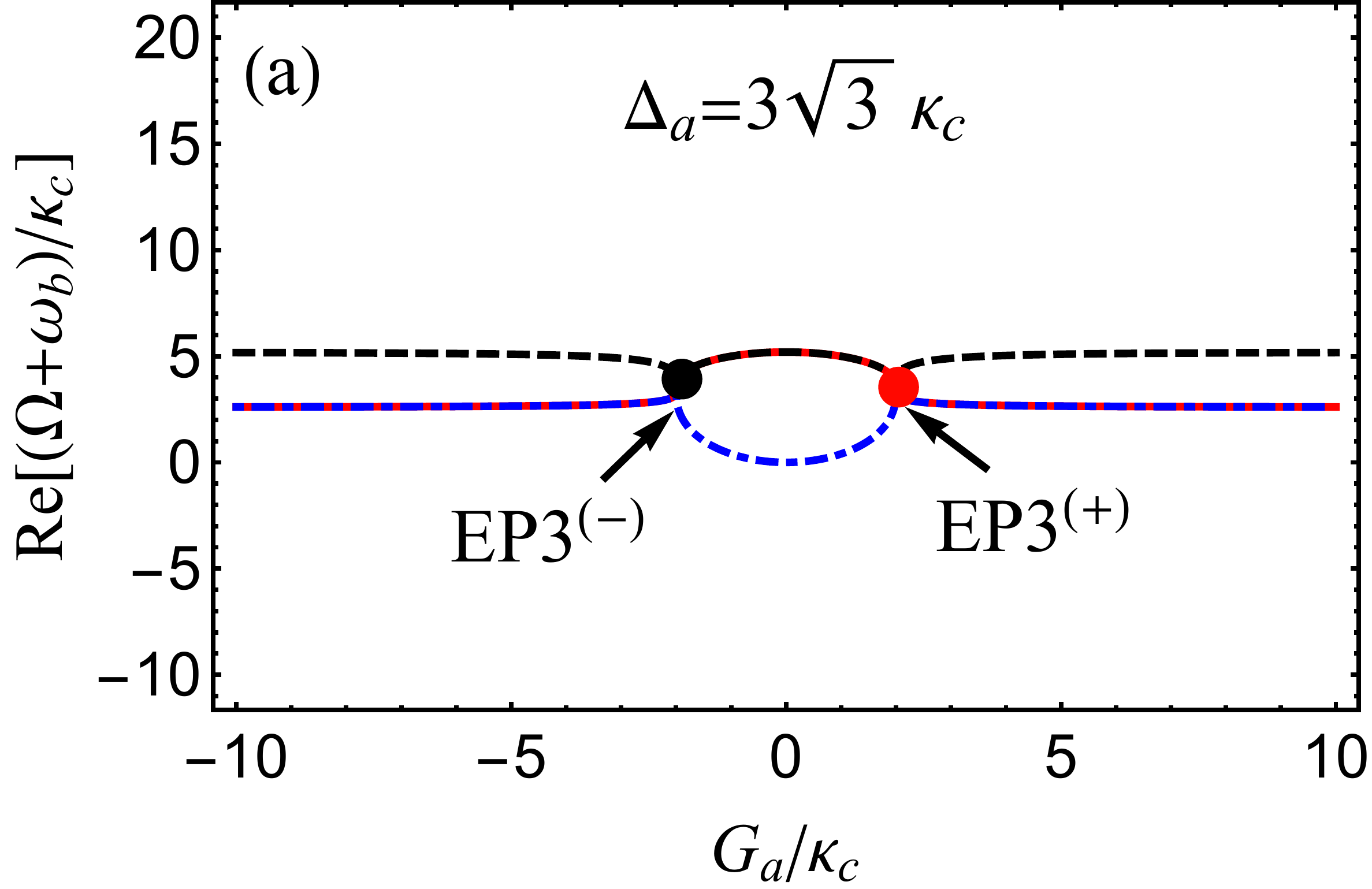}
	\includegraphics[scale=0.25]{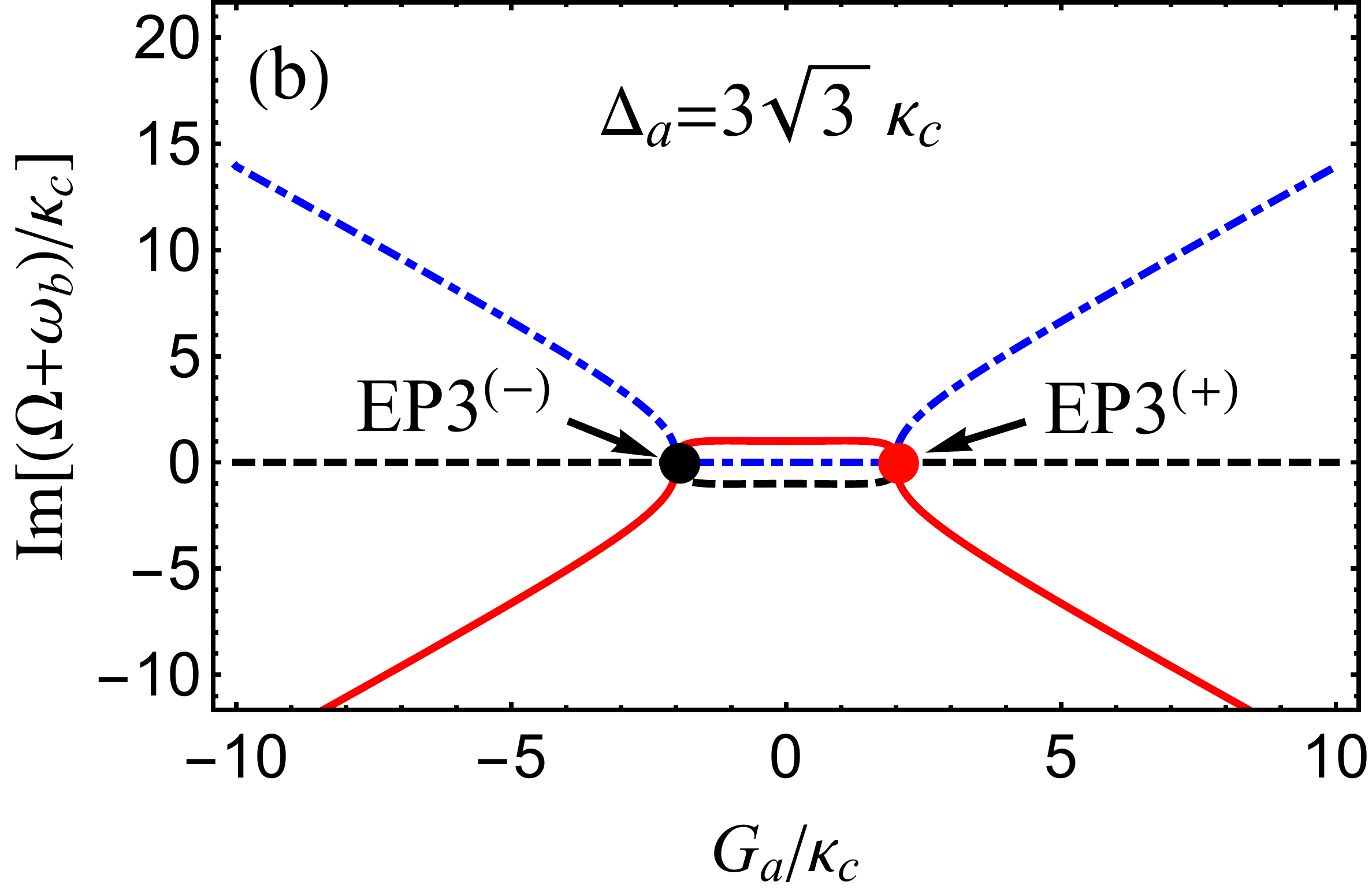}
	\includegraphics[scale=0.25]{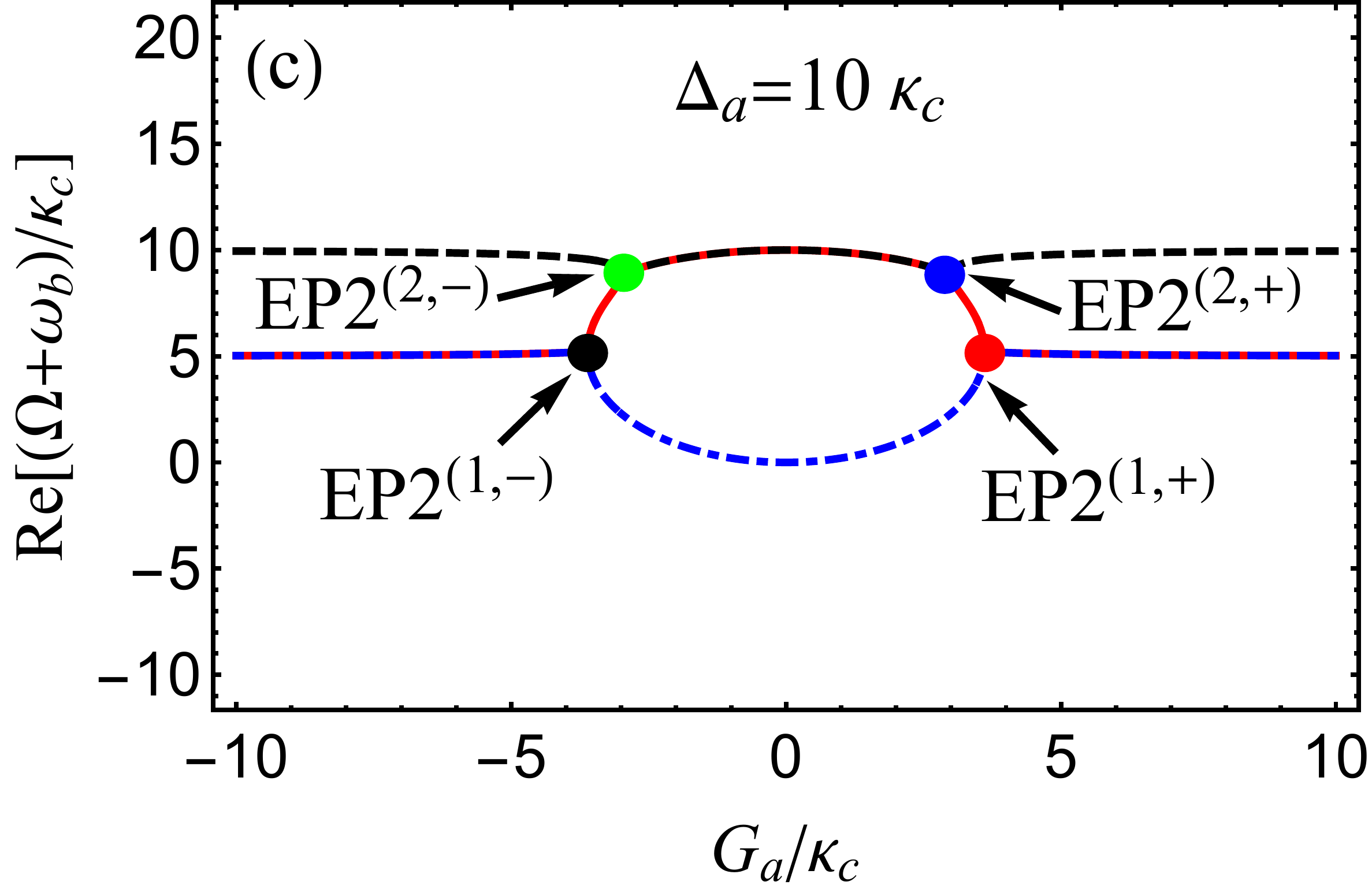}
	\includegraphics[scale=0.25]{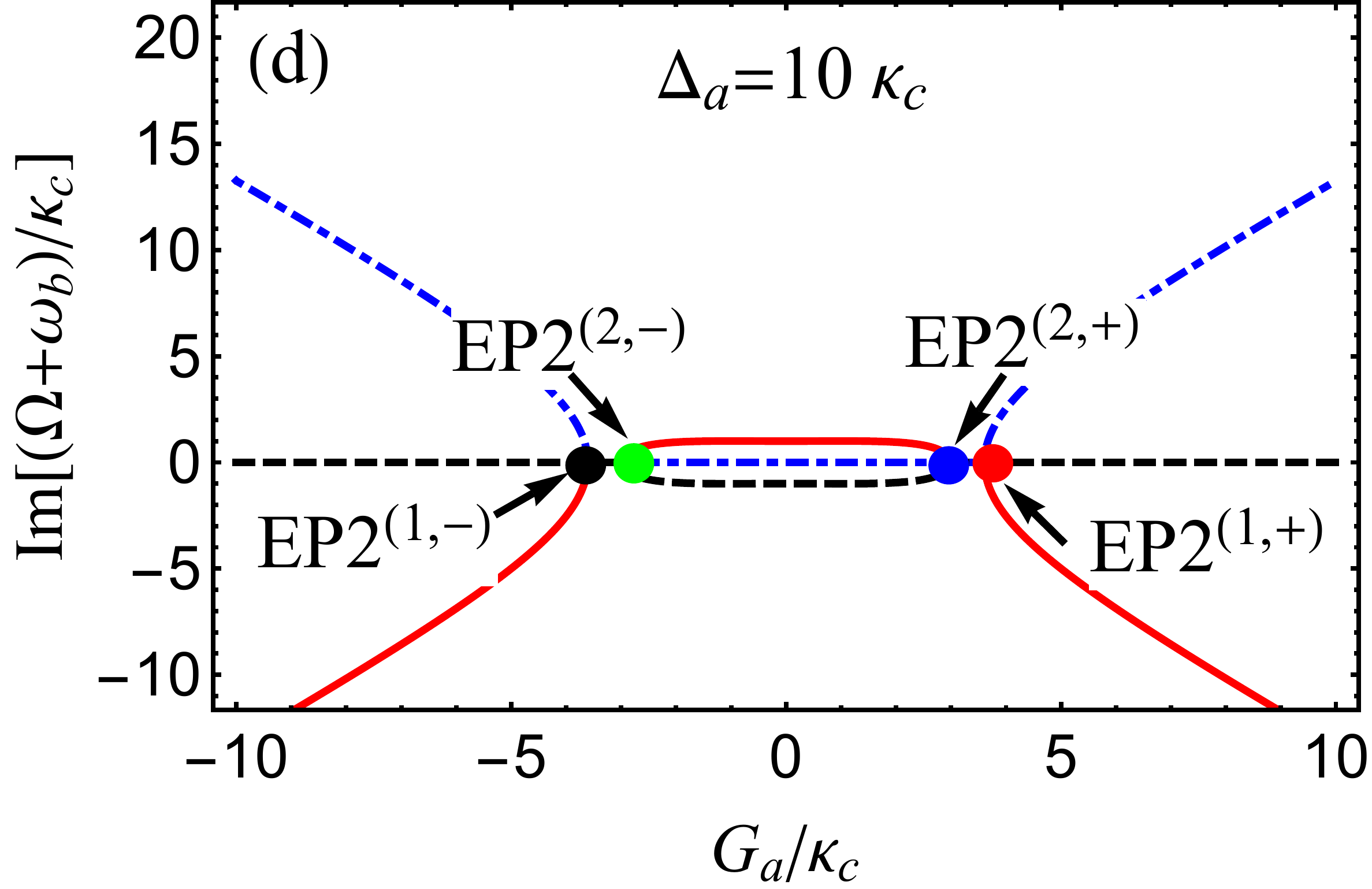}
	\caption{{The real and imaginary parts of the eigenvalue ($x=\Omega+\omega_b$) as a function of the normalized parameter $G_a/\kappa_c$ with (a) and (b) $\Delta_a=3\sqrt{3}\kappa_c$, (c) and (d) $\Delta_a=10\kappa_c$. Here $\eta=-1$, $\lambda=1$ and $\kappa_a+\kappa_b+\kappa_c=0.1\kappa_c$.}}\label{fig7}
\end{figure*}
 
\subsection{$\eta\neq-1$}

For the case of $\eta\neq-1$, we also have numerically proved that EP3 and EP2 can be predicted in our proposed blue-sideband optomechanical system by respectively taking $\eta=-1.1$ and $-0.8$ as examples in Figs.~\ref{fig2}(b) and \ref{fig2}(c). Here we further specifically study EP3 and EP2 by investigating the eigenvalues of $H_{\rm eff}$ with  $\eta=-1.1$ and $-0.8$. 

For $\eta=-1.1$ (or equivalently $\gamma_b>0$), which leads to $\lambda=\lambda_{\rm EP3}\approx1.33$ and $G_{\rm a, EP3}^{\rm min}=0.04\kappa_c$. we plot the real and imaginary parts of the eigenvalue ($x=\Omega+\omega_b$) vs the normalized parameter $G_a/\kappa_c$ with different $\lambda$ in Fig.~\ref{fig5}. From Figs.~\ref{fig5}(a) and \ref{fig5}(b) in which $\lambda=\lambda_{\rm EP3}$, we can see that the eigenvalue $x_3$ denoted by the black curve is always real for arbitrary $G_a\notin(-G_{\rm a, EP3}^{\rm min},G_{\rm a, EP3}^{\rm min})$, and the other two eigenvalues ($x_1$ and $x_2$) are a complex-conjugate pair except for at points $G_{\rm a, EP3}^{(\pm)}=\pm1.82\kappa_c$. At these two points, three eigenvalues coalesce into one value $\Omega_{\rm EP3}^{\pm}=3.7\kappa_c-\omega_b$, corresponding to two EP3s. When we take $\lambda=1.2\lambda_{\rm EP3}$, slightly derivating from $\lambda_{\rm EP3}$ at EP3s, the condition for predicting EP3s is broken, thus EP3 disappears. According to Fig.~\ref{fig2}(b), EP2 can be observed. In Figs.~\ref{fig5}(c) and \ref{fig5}(d), we plot the real and imaginary parts of the eigenvalue ($x=\Omega+\omega_b$) vs the normalized parameter $G_a/\kappa_c$ with $\lambda=1.2\lambda_{\rm EP3}$. It is not difficult to find that three eigenvalues are all real when $G_a>G_{\rm a, EP2}^{(+)}=2.44\kappa_c$ or $G_a<G_{\rm a, EP2}^{(-)}=-2.44\kappa_c$. At $G_a=G_{\rm a, EP2}^{(\pm)}$, $x_1$ and $x_3$ coalesce to $\Omega_{\rm EP2}^{\pm}=9.05\kappa_c-\omega_b$, corresponding to two EP2s. When $G_{\rm a, EP2}^{(-)}<G_a<G_{\rm a, EP2}^{(+)}$, the real parts of $x_1$ and $x_3$ are still degenerate, but their imaginary parts bifurcate into two values.

For the case of $\eta=-0.8$ (or equivalently $\gamma_b<0$), the behaviors of three eigenvalues  are similar to the case of $\eta=-1.1$ for predicting EP3s [see Figs.~\ref{fig6}(a) and \ref{fig6}(b)] and EP2s [see Figs.~\ref{fig6}(c) and \ref{fig6}(d)].

\section{Discussion and Conclusion}

{Note that our studies are constrainted in the pseudo-Hermitian condition, which can ensure the emergence of EPs in our proposed non-Hermitian COM system. But actually, the strict pseudo-Hermitian condition in general can not be fully satisfied, which means that the pseudo-Hermitian condition is broken [see Eq.~(\ref{eq20}) or Eq.~(\ref{phc})]. For example, the gain and loss in Eq.~(\ref{eq20}) are not balanced, i.e., $\kappa_a+\kappa_b+\kappa_c\neq0$. In this situation, we find that both EP3 and EP2 can also be predicted in our setup, as shown in Fig.~\ref{fig7} where $\eta=-1$ and $\kappa_a+\kappa_b+\kappa_c=0.1\kappa_c$ is taken. This shows that EPs in our proposal are robust against the slightly unbalanced gain and loss, which also reveals that the pseudo-Hermitian condition is neither sufficient nor necessary condition for predicting EPs. For the case of $\eta\neq-1$, we also numerically check it, and
the same result is obtained.}

In summary, we have proposed a blue-detuned non-Hermitian cavity optomechanical system consisting of a MR coupled to both a passive (loss) and an active (gain) cavities via radiation pressure for predicting EP3s. Under the pseudo-Hermitian condition,  the cases of the neural, loss and gain MRs are considered. By investigating the phase diagram of the discriminant, we find that both two degenerate or two non-degenerate EP3s can be predicted by tuning system parameters in the parameter space for the neutral MR. Also, four non-degenerate EP2s can be observed when system parameters {deviate} from EP3s, which is distinguished from the previous study in the red-detuned optomechanical system. For the gain (loss) MR, we find only two degenerate EP3s or EP2s can be predicted by tuning enhanced coupling strength. By studying the effect of parameters on EP3s or EP2s, we show that large parameters, such as frequency detuning and enhanced optomechanical coupling strength, can be employed to observe EPs more clearly. Our proposal is the first scheme to study higher-order EPs in the blue-detuned COM system, and it provides a potential way to investigate multimode quantum squeezing effects around higher-order EPs

\section*{ACKNOWLEDGMENTS}

This paper is supported by the key program of the Natural Science Foundation of Anhui (Grant No.~KJ2021A1301), the National Natural Science Foundation of China (Grants No.~12205069, No.~11904201 and No.~12104214), and the Natural Science Foundation of Hunan Province of China (Grant No. 2020JJ5466).

\setcounter{equation}{0}
\renewcommand{\theequation}{A\arabic{equation}}
\appendix{}

{\section{The derivation of the  effective Hamiltonian $H_{\rm eff}$}
In this Appendix, we derive the effective Hamiltonian given by Eq.~(\ref{eq16}) in the main text. Following the quantum Langevin equation method~\cite{Benguria-1981}, the dynamics of the proposed system including dissipations can be given by
\begin{align}\label{a1}
	\dot{a}=&-(\kappa_a+i\delta_a)a-ig_a a(b^\dag+b)+\Omega_a+\sqrt{2\kappa_a}a_{\rm in},\notag\\
	\dot{b}=&-(\gamma_b+i\omega_b)b-ig_a a^\dag a-ig_c c^\dag c+\sqrt{2\gamma_b}b_{\rm in},\\
	\dot{c}=&-(\kappa_c+i\delta_c)c-ig_c c(b^\dag+b)+\Omega_c+\sqrt{2\kappa_c}c_{\rm in},\notag
\end{align}
where $\kappa_{a(c)}$ is the decay rate of the cavity $a~(c)$, and $\gamma_b$ is the decay rate of the MR. {Note that when one of the cavities such as the cavity $a$ is subject to the dissipative gain, its corresponding dynamics in Eq.~(\ref{a1}) should be corrected as~\cite{Gardiner-2000} $ \dot{a}=-(i\delta_a-\kappa_a)a-ig_a a(b^\dag+b)+\Omega_a+\sqrt{2\kappa_a}a_{\rm in}$, which is different from the first equation in Eq.~(\ref{a1})}.  $a_{\rm in}$, $b_{\rm in}$ and $c_{\rm in}$ are vacuum input noises with zero expectation value, i.e., $\langle a_{\rm in}\rangle=\langle b_{\rm in}\rangle=\langle c_{\rm in}\rangle=0$. To linearize the nonlinear equations in Eq.~(\ref{a1}), we write the operators $a$, $b$, and $c$ as $a=a_s+\delta a,~b=b_s+\delta b,~c=c_s+\delta c$, where $a_s={\varepsilon_a}/{(\kappa_a+i\Delta_a)}$, $b_s=-i{(g_a  |a_s|^2+g_c |c_s|^2)}/{(\kappa_b+i\omega_b)}$, $c_s={\varepsilon_c}/{(\kappa_c+i\Delta_c)}$ are steady-state values, and $\delta a$, $\delta b$, $\delta c$ are fluctuation operators. Then we substitute these transformations into Eq.~(\ref{a1}). In the strong-field limit, i.e., $|a_s|,|c_s|\gg1$, higher-order fluctuation terms can be safely neglected. Thus, the dynamics of the fluctuation operators in Eq.~(\ref{a1}) can be linearized as
\begin{align}\label{a2}
	\dot{\delta a}=&-(\kappa_a+i\delta_a^\prime)\delta a-iG_a(\delta b^\dag+\delta b)+\sqrt{2\kappa_a}a_{\rm in},\notag\\
	\dot{\delta b}=&-(\gamma_b+i\omega_b)\delta b-i (G_a^* \delta a+G_a\delta a^\dag)\notag\\
	&-i (G_c^* \delta c+G_c\delta c^\dag)+\sqrt{2\gamma_b}b_{\rm in},\\
	\dot{\delta c}=&-(\kappa_c+i\delta_c^\prime)\delta c-iG_c(\delta b^\dag+\delta b)+\sqrt{2\kappa_c}c_{\rm in}.\notag
\end{align}
where $	\delta_a^\prime=\delta_a+g_a a_s(b_s^*+b_s)$ and $	\delta_c^\prime=\delta_c+g_c c_s(b_s^*+b_s)$ are the effective frequency detunings of the cavity $a$ and the cavity $c$, respectively, shifted by the displacement of the MR.  In general, such frequency shifts are tiny due to weak single-photon optomechanical coupling strengths. Experimentally, $\delta_{a(c)}^\prime\approx\delta_{a(c)}$ are used. $G_a=g_a a_s$ and $G_c=g_c c_s$ are the effective enhanced optomechanical coupling strengths, which can be tuned by the amplitudes of the two laser fields. Under the condition $|\delta_{a(c)}^\prime+\omega_b|\ll|\delta_{a(c)}^\prime-\omega_b|$ and $|G_{a(c)}|\ll|\delta_{a(c)}^\prime|$, the fast oscillating terms in Eq.~(\ref{a2}) can be neglected, then Eq.~(\ref{a2}) reduces to
\begin{align}\label{a3}
	\dot{\delta a}=&-(\kappa_a+i\delta_a^\prime)\delta a-iG_a\delta b^\dag+\sqrt{2\kappa_a}a_{\rm in},\notag\\
	\dot{\delta b}=&-(\gamma_b+i\omega_b)\delta b-i G_a\delta a^\dag
	-i G_c\delta c^\dag+\sqrt{2\gamma_b}b_{\rm in},\notag\\
	\dot{\delta c}=&-(\kappa_c+i\delta_c^\prime)\delta c-iG_c\delta b^\dag+\sqrt{2\kappa_c}c_{\rm in}.
\end{align}
By rewriting the equations of motion in Eq.~(\ref{a3}) as $\dot{\delta a}=-i[\delta a,H_{\rm eff}]+\sqrt{2\kappa_a}a_{\rm in}$, $\dot{\delta b}=-i[\delta b,H_{\rm eff}]+\sqrt{2\gamma_b}b_{\rm in}$, and $\dot{\delta c}=-i[\delta c,H_{\rm eff}]+\sqrt{2\kappa_c}c_{\rm in}$,  the effective non-Hermitian Hamiltonian in the blue-sideband regime can be obtained,
\begin{align}
	H_{\rm eff}=&(\delta_a^\prime-i\kappa_a) \delta a^\dag  \delta a+(\omega_b-i\gamma_b) \delta b^\dag \delta b+(\delta_c^\prime-i\kappa_c) \delta c^\dag \delta c\notag\\
	&+G_a (\delta a^\dag \delta b^\dag+\delta a \delta b)+G_c(\delta b^\dag\delta c^\dag +\delta b\delta c ),\label{a4}
\end{align}
which is just the effective Hamitonian in Eq.~(\ref{eq16}).}

{\section{Stability}}
\renewcommand{\theequation}{B\arabic{equation}}	

{From Eq.~(\ref{a3}), we can obtain the following equations,
\begin{align}\label{a5}
	\dot{\delta a}=&-(\kappa_a+i\delta_a^\prime)\delta a-iG_a\delta b^\dag+\sqrt{2\kappa_a}a_{\rm in},\notag\\
	\dot{\delta b}=&-(\gamma_b+i\omega_b)\delta b-i G_a\delta a^\dag
	-i G_c\delta c^\dag+\sqrt{2\gamma_b}b_{\rm in},\notag\\
	\dot{\delta c}=&-(\kappa_c+i\delta_c^\prime)\delta c-iG_c\delta b^\dag+\sqrt{2\kappa_c}c_{\rm in}.\\
	\dot{\delta a^\dag}=&-(\kappa_a-i\delta_a^\prime)\delta a^\dag+iG_a\delta b+\sqrt{2\kappa_a}a_{\rm in}^\dag,\notag\\
	\dot{\delta b^\dag}=&-(\gamma_b-i\omega_b)\delta b^\dag+i G_a\delta a
	+i G_c\delta c+\sqrt{2\gamma_b}b_{\rm in}^\dag,\notag\\
	\dot{\delta c^\dag}=&-(\kappa_c-i\delta_c^\prime)\delta c^\dag+iG_c\delta b+\sqrt{2\kappa_c}c_{\rm in}^\dag.\notag
\end{align}
By setting 
\begin{align}\label{a6}	
    \delta X_a=& \frac{a+a^\dag}{\sqrt{2}},~~\delta Y_a= \frac{a-a^\dag}{i\sqrt{2}},\notag\\
	\delta X_b=& \frac{b+b^\dag}{\sqrt{2}},~~\delta Y_b= \frac{b-b^\dag}{i\sqrt{2}},\notag\\
	\delta X_c=& \frac{c+c^\dag}{\sqrt{2}},~~\delta Y_c= \frac{c-c^\dag}{i\sqrt{2}},\notag\\
    \delta X_{a_{\rm in}}=& \frac{a_{\rm in}+a_{\rm in}^\dag}{\sqrt{2}},~~\delta Y_{a_{\rm in}}= \frac{a_{\rm in}-a_{\rm in}^\dag}{i\sqrt{2}},\notag\\
    \delta X_{b_{\rm in}}=& \frac{b_{\rm in}+b_{\rm in}^\dag}{\sqrt{2}},~~\delta Y_{b_{\rm in}}= \frac{b_{\rm in}-b_{\rm in}^\dag}{i\sqrt{2}},\notag\\
    \delta X_{c_{\rm in}}=& \frac{c_{\rm in}+c_{\rm in}^\dag}{\sqrt{2}},~~\delta Y_{c_{\rm in}}= \frac{c_{\rm in}-c_{\rm in}^\dag}{i\sqrt{2}},
\end{align}
Eq.~(\ref{a5}) can be rewritten as
\begin{align}
\dot{u}=Mu+f_{\rm in},	
\end{align}
where $u=[\delta X_a,\delta Y_a, \delta X_b,\delta Y_b,\delta X_c,\delta Y_c]^T$, $f_{\rm in}=[\delta X_{a_{\rm in}},\delta Y_{a_{\rm in}}, \delta X_{b_{\rm in}}, \delta Y_{b_{\rm in}},\delta X_{c_{\rm in}},\delta Y_{c_{\rm in}}]^T$, and $M$ is given  by
\begin{equation}
	M=\left(
	\begin{array}{cccccc}
		-\kappa_a& \delta_a^\prime& 0& -G_a & 0&0\\
		-\delta_a^\prime & -\kappa_a& -G_a&0 &  0&0\\
		0 &  -G_a  & -\kappa_b &\omega_b&0&-G_c\\
		-G_a &0 &-\omega_b& -\kappa_b &-G_c&0 \\
		0&0&0&-G_c&-\kappa_c&\delta_c^\prime\\
		0&0&-G_c&0&-\delta_c^\prime&-\kappa_c\\
	\end{array}
	\right).
\end{equation}
The considered COM system is stable only when the real parts of the eigenvalues $\lambda$ of the matrix $M$ are all negative, which can be judged by the Routh-Hurwitz criterion~\cite{Gradshteyn}. To using this criterion, we expand the characteristic equation $|M-I\lambda|=0$ as $\lambda^6+c_5 \lambda^5+c_4\lambda^4+c_3\lambda^3+c_2\lambda^2+c_1\lambda+c_0=0$, where the coefficients $c_j$ with $j=0,1,...,5$ can be derived using straightforward but tedious algebra. Interestingly, we find $c_5=0$ when $\kappa_a+\gamma_b+\kappa_c=0$, which breaks the Routh-Hurwitz criterion for prediction of the stability. This indicates that when the pseudo-Hermitian condition is strictly satisfied, the considered system is possibly unstable. To ensure the system stable, $\kappa_a+\gamma_b+\kappa_c>0$ is required in experiment. This requirement can be achieved when two cavities are gain-loss balanced and the loss mechanical resonatr is employed.  Although the pseudo-Hermitian condition is broken, EP3s or EP2s can also be predicted (see Fig.~\ref{fig7}).  Other stable conditions obtained from the Routh-Hurwitz criterion can be well satisfied. This is due to the tunable frequency detunings and linearized optomechanical coupling strengths via tuning driving fields.}

\end{document}